\documentclass[prd,superscriptaddress,amsfonts,amssymb,amsmath,showpacs,twocolumn]{revtex4-2}
\usepackage{bm}
\usepackage{amsfonts}
\usepackage{latexsym}
\usepackage[latin1]{inputenc}
\usepackage{graphicx}
\usepackage{amsmath,eqnarray}
\usepackage{palatino}
\usepackage{mathpazo}
\usepackage{textcomp}
\usepackage{float}
\usepackage{booktabs}
\usepackage{dcolumn}
\usepackage{hyperref}
\usepackage{mathtools}
\hypersetup{colorlinks,citecolor=red}

\usepackage{amsmath}
\usepackage{xcolor}
\usepackage{orcidlink}
\usepackage[caption=false]{subfig}
\usepackage{commath}
\def\jnl@style{\it}
\def\aaref@jnl#1{{\jnl@style#1}}

\def\aaref@jnl#1{{\jnl@style#1}}

\def\aj{\aaref@jnl{AJ}}                   
\def\apj{\aaref@jnl{ApJ}}                 
\def\apjl{\aaref@jnl{ApJ}}                
\def\apjs{\aaref@jnl{ApJS}}               
\def\apss{\aaref@jnl{Ap\&SS}}             
\def\aap{\aaref@jnl{A\&A}}                
\def\aapr{\aaref@jnl{A\&A~Rev.}}          
\def\aaps{\aaref@jnl{A\&AS}}              
\def\mnras{\aaref@jnl{Mon.~Not.~Roy.~Astron.~Soc.}}             
\def\prd{\aaref@jnl{Phys.~Rev.~D}}        
\def\prc{\aaref@jnl{Phys.~Rev.~C}}  
\def\prl{\aaref@jnl{Phys.~Rev.~Lett.}}    
\def\qjras{\aaref@jnl{QJRAS}}             
\def\skytel{\aaref@jnl{S\&T}}             
\def\ssr{\aaref@jnl{Space~Sci.~Rev.}}     
\def\zap{\aaref@jnl{ZAp}}                 
\def\nat{\aaref@jnl{Nature}}              
\def\aplett{\aaref@jnl{Astrophys.~Lett.}} 
\def\apspr{\aaref@jnl{Astrophys.~Space~Phys.~Res.}} 
\def\physrep{\aaref@jnl{Phys.~Rep.}}      
\def\physscr{\aaref@jnl{Phys.~Scr}}       
\def\commat{\aaref@jnl{Comm.~Math.~Phys.}}              
\def\science{\aaref@jnl{Science}}               
\def\cqg{\aaref@jnl{Classical Quant.~Grav.}}            
\def\jpcs{\aaref@jnl{JPCS}}                                     
\def\ijmpd{\aaref@jnl{Int.~J.~Mod.~Phys.~D}}                    
\def\grg{\aaref@jnl{Gen.~Relat.~Gravit.}}               
\def\rpp{\aaref@jnl{Rep.~Prog.~Phys.}}          
\def\npa{\aaref@jnl{Nucl.~Phys.~A}}        
\def\lrr{\aaref@jnl{Living Rev.~Rel.}}                   
\def\jcap{\aaref@jnl{J.~Cosmology Astropart.~Phys.}}    
\def\rmp{\aaref@jnl{Rev.~Mod.~Phys.}}   
\def\epjc{\aaref@jnl{Eur.~Phys.~J.~C}}

\allowdisplaybreaks[1]

\begin{document}

\color{black}

\title{Geometrically deformed charged anisotropic models in $f(Q,T)$ gravity}

\author{Sneha Pradhan\orcidlink{0000-0002-3223-4085}}
\email{snehapradhan2211@gmail.com}
\affiliation{Department of Mathematics, Birla Institute of Technology and
Science-Pilani,\\ Hyderabad Campus, Hyderabad-500078, India.}

\author{Sunil Kumar Maurya\orcidlink{0000-0003-4089-3651}}
\email{sunil@unizwa.edu.om}
\affiliation{Department of Mathematical and Physical Sciences,\\ College of Arts and Sciences, University of Nizwa, Nizwa 616, Sultanate of Oman}

\author{Pradyumn Kumar Sahoo\orcidlink{0000-0003-2130-8832}}
\email{pksahoo@hyderabad.bits-pilani.ac.in}
\affiliation{Department of Mathematics, Birla Institute of Technology and
Science-Pilani,\\ Hyderabad Campus, Hyderabad-500078, India.}

\author{Ghulam Mustafa \orcidlink{0000-0003-1409-2009}}
\email{gmustafa3828@gmail.com}
\affiliation{Department of Physics, Zhejiang Normal University, Jinhua 321004, People's Republic of China}

\date{\today}

\begin{abstract}
In this study, we developed the geometrically deformed compact objects in the $f(Q, T)$ gravity theory under an electric field through gravitational decoupling via. minimal geometric deformation (MGD) technique for the first time.  The decoupled field equations are solved via two different mimic approaches $\theta_0^0 = \rho$ and $\theta_1^1 = p_r$ through the Karmarkar condition. We conduct physical viability tests on our models and examine how decoupling parameters affect the physical qualities of objects. The obtained models are compared with the observational constraints for neutron stars PSR J1810+174, PSR J1959+2048, and PSR J2215+5135, including GW190814.  Particularly, by modifying parameters $\alpha$ and $n$, we accomplish the occurrence of a "\textit{mass gap}" component. The resulting models exhibit stable, well-behaved mass profiles, regular behaviour, and no gravitational collapse, as verified by the Buchdahl--Andr\'{e}asson's limit. Furthermore, we provide a thorough physical analysis that is based on two parameters: $n$ ($f(Q,T)$--coupling parameter) and $\alpha$ (decoupling parameter). This work extends our current understanding of compact star configurations and sheds light on the behaviour of compact objects in the $f(Q,T)$ gravity.  \\


\end{abstract}

\maketitle

\section{Introduction}\label{sec:1}
The study of our expanding Universe provides us with valuable perspectives on its formation and mysterious properties. In recent years, various astrophysical phenomena, including the origin and evolution of cosmic structures, have attracted significant interest among researchers. Astrophysical objects are often regarded as the fundamental building blocks of galaxies, which are arranged in a structured manner inside a cosmic web. The phenomenon of stellar collapse, which occurs as a consequence of gravitational forces acting inward, leads to the creation of novel compact objects. To investigate the internal geometry of these objects, it is essential to get analytical solutions for the non-linear field equations. Despite the fundamental non-linearity of these partial differential equations, numerous researchers have successfully formulated precise and feasible astrophysical and cosmological solutions. The initial solution of the Einstein field equations for a spherically symmetric object in a vacuum was formulated by Schwarzschild \cite{SC}.

Stars have the potential to be regarded as self-gravitating fluid spheres. An analytical solution to the aforementioned type of fluid was provided by Krori and Barua in their work \cite{KB}. When considering isotropic stresses, it has been observed that the static spheres in general relativity (GR) composed of ordinary fluid adhere to the Buchdahl limit on compactness, which states that $\frac{2M}{R}<\frac{8}{9}$ \cite{Buchdahl1} while this limit changes in the presence of electric charge~\cite{HA}. This anisotropy issues can appear in the system under diverse circumstances :
In dense environments \cite{MR, VC},  when electromagnetic or fermionic factors are involved, or in configurations of pion condensation within neutron stars, solid cores\cite{RF}, superfluidity\cite{BC}, etc. Anisotropy is widespread and can even be seen in everyday objects like a soap bubble that exhibits anisotropic stress \cite{JG}.
The investigation of anisotropic stars in the context of GR has primarily focused on static, spherically symmetric solutions \cite{RL, PS}.  

In the study \cite{MK}, scientists introduced a category of precise solutions to Einstein's gravitational field equations by assuming a specific form for the anisotropy factor. According to their numerical findings, the model's fundamental physical characteristics (mass and radius) can accurately characterize various astrophysical objects, like neutron stars. In the work \cite{SKM1}, the researchers have applied the embedding approach under the most straightforward linear function of the matter-geometry coupling to study the anisotropic strange star. The author in \cite{SB} did a study on an anisotropic spherically symmetric strange star, employing the Tolman-Kuchowicz type's metric potentials in their analysis. The study \cite{SK2} presented a novel, precise, and analytical solution to the Einstein-Maxwell field equations, which accurately describe compact, anisotropic, and charged stellar objects. The solution is smoothly matched with the external Reisner-Nordstrom spacetime to determine the system's physical characteristics. Over the past ten years, numerous studies published in the literature, have included ideas about using anisotropic matter distribution to solve analytical models for Einstein field equations \cite{1,2,3,4,5,6,7,8,9,10,11,12,13,14,15}. 

One of the pillars of GR is the curvature that originates from Riemannian geometry and is denoted by the Ricci scalar $R$. The modified $f(R)$ gravity, a fundamental modification of GR, replaces the Ricci scalar R with some general function of $R$ \cite{JB1, JB2, JB3}. Moreover, in non-Riemannian geometry, other alternatives exist to GR, one of which is the teleparallel equivalent of GR (TEGR). In TEGR, the gravitational interactions are described using the torsion ($T$) known as $f(T)$ gravity. Another alternative is symmetric teleparallel gravity or $f(Q)$ gravity, where the non-metricity $Q$ represents a building block of the gravitational field between the particles. The "coincidencident gauge," known in this theory, is often used to ensure that the affine connection is zero while the metric remains the primary fundamental variable. This specific gauge choice has been consistently utilized in various research endeavors exploring extensions of the Standard Theory of General Relativity (STGR). The $f(Q)$ theory gained popularity among cosmologists and was introduced by J. B. Jimenez and colleagues in 2018 \cite{JM}. In $f(Q)$ gravity, the key distinction from classical GR lies in the affine connection rather than the properties of the physical manifold itself. In the work \cite{JM}, the authors demonstrated that $f(Q)$ gravity is equivalent to GR in flat space. Additionally, Lin and Zhai \cite{LZ} studied the application of $f(Q)$ gravity to spherically symmetric configurations, examining both exterior and interior solutions of compact stars. Their work shed light on the consequences of employing $f(Q)$ gravity in this context. In a different study \cite{PB}, researchers delved into the properties of static anisotropic hybrid stars, which consist of strange quark matter (SQM) and ordinary baryonic matter (OBM) distributions. Moreover, several works on spherically symmetric compact objects with an anisotropic fluid distribution have been explored in a broader range \cite{SM, SP, YK1, YK2}. Subsequently, a further extension of $f(Q)$ gravity called $f(Q, T)$ gravity was introduced by Yixin and colleagues \cite{YX}. This theory proposes a connection between the gravitational effects and two key factors: the non-metricity function $Q$ and the trace of the energy-momentum tensor $T$. This framework suggests that gravitational interactions are linked to the non-metricity function $Q$ and influenced by manifestations originating from the quantum field due to the presence of the energy-momentum tensor $T$. Although the proposal of this notion of gravity is relatively recent, significant advancements have been achieved in its study. Numerous studies have been conducted to explore its theoretical aspects \cite{Najera, Bhattacharjee} and its observational dimensions \cite{Arora1}. The investigation conducted by Harko et al. \cite{T.Harko} focused on exploring novel associations between non-metricity and matter.\\
Extracting an anisotropic solution of the non-linear system of field equations presents a challenging problem, mostly due to the difference between the number of unknowns and the number of equations. Various methods have been implemented to address this issue and facilitate the development of plausible solutions. The approach of gravitational decoupling using MGD, as developed by Ovalle \cite{JO1,JO2}, is employed to identify novel solutions corresponding to various relativistic distributions in astrophysics. This method modifies the radial metric component and produces two distinct systems of differential equations derived from the field equation system. One system includes the seed source, and the impact of an extra source influences the other system. The two sets are solved individually, and the overall solution of the system is derived by applying the superposition principle. The MGD approach effectively inhibits energy transfer between matter sources, maintaining the self-gravitating system's spherical symmetry.
In accordance with the MGD approach, the solution proposed by Ovalle and Linares~\cite{JO3} was examined in the context of a braneworld scenario. It was determined that the compactness factor experiences a reduction due to the influence of fluid dispersion in the bulk. In a further study conducted in 2018, Ovalle and his colleagues \cite{JO4} proposed an anisotropic interior solution of perfect fluid distribution. This solution was achieved by incorporating an additional gravitational source. Gabbanelli et al. \cite{LG} examined the prominent characteristics of the anisotropic extension of the Durgapal-Fuloria model through gravitational decoupling. In their study, Sharif and Sadiq \cite{MS} employed this methodology to construct anisotropic models with charge distribution and investigated celestial objects' physical characteristics. In the study \cite{Sk1}, the author extends the Maurya-Gupta isotropic fluid solution \cite{Mu1} to the Einstein field equations in an anisotropic domain. To do this, they have implemented the technique of gravitational decoupling through the method of minimal geometric distortion for the strange star candidate LMC X-4. In addition to the ones mentioned above, several literary works explore the concept of gravitational decoupling using the MGD technique to explore celestial objects\cite{a,b,c,d,e,f}.
Researchers have delved into the gravitational decoupling method to investigate not only exotic celestial bodies like strange stars but also to study black holes and wormhole objects, as evidenced by references \cite{hw1,hw2,hw3,hw4}. In this work, we have studied the geometrically deformed compact star under the influence of an electric field through the gravitational decoupling method. The stability of a star is primarily due to the delicate equilibrium between the inward gravitational force and the outward pressure generated by nuclear fusion reactions in its core. Gravitational attraction tends to compress the star, while the energy and radiation produced by the fusion create pressure that counteracts this compression. This balance, known as hydrostatic equilibrium, ensures the star remains stable. But, when there is inadequate pressure to counterbalance the gravitational pull within a stellar entity, it experiences abrupt gravitational collapse, resulting in significant alterations to its physical properties. These observations have led researchers to postulate that the presence of charge in compact objects might inhibit gravitational attraction, with the repulsive Coulomb force supplementing the pressure gradient, as indicated by \cite{Day}. P. M. Takis et al. \cite{charge1} has shown the influence of electric charge in the anisotropic compact object with the conformal symmetry. Apart from these, some excellect work on a charged compact stars can be found in the literatur \cite{charge2,charge3}.\\
Our paper is organized as follows: 
In Sec.-\ref{II}, the basic mathematical configuration of $f(Q, T)$ gravity has been given briefly. In Sec.-\ref{III}, we have split two sets of equations by introducing the MGD technique. In Sec.-\ref{IV}, the matching condition between the interior and exterior space-time has been described. In Sec.-\ref{VI}, we have derived an exact solution for the first system by introducing the embedding Class-I method in Sec.-\ref{V}. Next, we have solved the second set of equations by mimicking the pressure and density constraints in Sec.-\ref{VII}. After that, the physical analysis has been done in Sec.-\ref{VIII}. Furthermore, in Sec.-\ref{XI}, we have discussed the stability of our suggested decoupled strange star model in $f(Q, T)$ gravity. At last, in Sec.-\ref{sec:XA}, we have explored the maximum allowable mass for a strange star and verified with the observational findings, and in Sec.-\ref{XII}, the result and discussion are presented.

\section{A brief review of $f(Q,T)$ gravity}\label{II}
The generalized action integral for $f(Q,T)$ gravity can be written as \cite{YX},
\begin{eqnarray}
    \label{1}
&&\hspace{-0.5cm} S=\int \sqrt{-g}\Big[\frac{1}{16 \pi }f(Q,T)\nonumber\\&&\hspace{0.5cm}+\mathcal{L}_M+\alpha \mathcal{L}_\Theta\Big] d^4x.
\end{eqnarray}

Where $\mathcal{L}_M$ stands for the Lagrangian density for matter fields related to the stress energy-momentum tensor ($T_{\mu\eta}$), $\mathcal{L}_{\Theta}$ represents the Lagrangian density for the new gravitational sector which can be called as ``$\Theta$ gravitational sector"($\Theta_{\mu\eta}$). This new additional contribution is always able to provide the modifications of matter fields in $f(Q,T)$ gravity, and it can be combined as a component of the effective energy-momentum tensor $T^{\text{eff}}_{\mu\eta}=T_{\mu\eta}+\alpha\,\Theta_{\mu\eta}$. Here $\alpha$ represents the coupling constant between the matter fields and $\Theta$ gravitational sector and $g$ is the determinant of the metric tensor.

The expression for non-metricity tensor concerning the affine connection is given by,
\begin{eqnarray}
\label{4}
Q_{\kappa\mu\eta}\equiv \nabla_\kappa g_{\mu\eta}=\partial_{\kappa}\,g_{\mu\eta}-\Gamma^{\delta}{}_{\kappa\mu}\,g_{\delta \eta}-\Gamma^{\delta}{}_{\kappa\eta}\,g_{\mu\delta}.
\end{eqnarray}
Here $\Gamma^{\delta}{}_{\kappa\mu}$ represents the usual affine connection whose form is given by,

\begin{eqnarray}
    \Gamma^{\delta}{}_{\kappa\mu} = \left\{
\begin{array}{c}
\delta \\ \kappa \mu 
\end{array}
\right\}+K^{\delta}{}_{\kappa \mu}+L^{\delta}{}_{\kappa\mu}.
\end{eqnarray}
Where $\left\{
\begin{array}{c}
\delta \\ \kappa \mu 
\end{array}
\right\}$, $L^{\delta}{}_{\kappa\mu}$, and $K^{\delta}{}_{\kappa \mu}$ are known as Levi-Civita connection, contorsion tensor and disformation tensor respectively. The mathematical expression for the above terms are given by,
\begin{eqnarray}
    \left\{
\begin{array}{c}
\delta \\ \kappa \mu 
\end{array}
\right\} &=& \frac{1}{2}\,g^{\delta\varphi}\left(\partial_{\kappa}g_{\varphi\mu}+\partial_{\mu}g_{\delta\eta}-\partial_{\delta}g_{\kappa\mu}\right),\\
K^{\delta}{}_{\kappa \mu} &=& \frac{1}{2} T^{\delta}{}_{\kappa \mu} + T_(\kappa{\,}^{\delta}{\,}_\mu),\\
L^{\delta}{}_{\kappa\mu} &=& \frac{1}{2} Q^{\delta}{}_{\kappa\,\mu}-Q_(\kappa{\,}^{\delta}{\,}_\mu).
\end{eqnarray}
Where $T^{\delta}{}_{\kappa \mu}$ is known as torsion tensor. Another important quantity in this STEGR formalism is the superpotential $P_{\,\,\,\,\mu\eta}^{\varphi}$ is defined as,

\begin{eqnarray}
\label{5}
P_{\,\,\,\,\mu\eta}^{\varphi}=-\frac{1}{2}L^\varphi_{\,\,\,\,\mu\eta}+\frac{1}{4}\left(Q^{\varphi}-\tilde{Q^{\varphi}}\right)g_{\mu\eta}-\frac{1}{4}\delta^{\varphi}_{\,\,(\mu\,} Q_{\eta)},
\end{eqnarray}

where the trace of the non-metricity tensor is given by,

\begin{equation*}
Q_{\varphi}=Q_{\varphi\,\,\,\,\,\mu}^{\,\,\,\mu}, \quad \quad \tilde{Q}_{\varphi}=Q^{\mu}_{\,\,\,\,\varphi\mu}.
\end{equation*}

Finally, the non-metricity scalar is defined as,
\begin{eqnarray}
\label{6}
Q=-Q_{\varphi\mu\eta}P^{\varphi\mu\eta}.
\end{eqnarray}
 The field equations of $f(Q, T)$ theory by varying the action \eqref{1} with respect to the metric tensor inverse $g^{\mu\eta}$ is obtained as
\begin{eqnarray}
\label{field}
&&\frac{2}{\sqrt{-g}}\nabla_{\varphi}\left(f_{Q}\sqrt{-g}\,P^{\varphi}_{\,\,\,\,\mu\eta}\right)+f_{Q}\left(P_{\mu\varphi\kappa}Q_{\eta}^{\,\,\,\varphi\kappa}-2Q^{\varphi\kappa}_{\,\,\,\,\,\,\,\,\mu}\, P_{\varphi\kappa\eta}\right)\nonumber \\&&\hspace{0.9cm}+\frac{1}{2}f\,g_{\mu\eta}=-8\pi T_{\mu\eta}^{\text{eff}}+f_{T}\left(T_{\mu\eta}+\Phi_{\mu\eta}\right).
\end{eqnarray}

Where, $f_Q = \frac{\partial f(Q,T)}{\partial Q}$, $f_T =\frac{\partial f(Q,T)}{\partial T}$ and $ T_{\mu\eta}^{eff} = T_{\mu\eta}+\,\alpha \, \Theta_{\mu\eta}$. Here the usual energy momentum tensor $T_{\mu\eta}$ and the additional source $\Theta_{\mu\eta}$ is defined as,
\begin{eqnarray}
 T_{\mu\eta} &=& -\frac{2}{\sqrt{-g}}\frac{\delta(\sqrt{-g}\mathcal{L}_m)}{\delta g^{\mu\eta}}, \\  \Theta_{\mu\eta} &=& -\frac{2}{\sqrt{-g}}\frac{\delta(\sqrt{-g}\mathcal{L}_{\Theta})}{\delta g^{\mu\eta}}.
\end{eqnarray}
In the Eq.~(\ref{field}), another unknown quantity is the hyper-momentum tensor which is denoted as $\Phi_{\mu\eta}$ and defined by $\Phi_{\mu\eta} = g^{\varphi \beta}\frac{\delta T_{\varphi \beta}}{\delta g^{\mu\eta}}$.\\
Furthermore, by utilizing Eq.~(\ref{1}), we can deduce an additional constraint given by,
\begin{eqnarray}
    \nabla_{\mu}\nabla_{\eta}(\sqrt{-g}f_{Q}P^{\varphi}{}_{\mu\eta})=0.
\end{eqnarray}
The affine connection is made possible by the curvature-free and torsion-free constraints as \cite{JB2}
\begin{eqnarray}\label{eq:13}
    \Gamma^{\varphi}{}_{\mu\eta}=\left(\frac{\partial x^{\varphi}}{\partial \xi^{\alpha}}\right) \partial_{\mu}\partial_{\eta} \xi^{\alpha}.
\end{eqnarray}
Now we can make a special coordinate choice, the so called co-incident gauge, so that the affine connection $\Gamma^{\varphi}{}_{\mu\eta}=0$. Then, the non-metricity equation reduces to
\begin{eqnarray}
    Q_{\kappa\mu\eta}\equiv \nabla_\kappa g_{\mu\eta}=\partial_{\kappa}\,g_{\mu\eta}.
\end{eqnarray}

Consequently, this simplifies the computation process since the metric is the primary variable now. Nevertheless, it should be noted that the action no longer maintains diffeomorphism invariance, except with the STGR \cite{JB3}. Such difficulty can be overcome by employing the covariant formulation of $f(Q)$ gravity. Given that the affine connection mentioned in Equation \eqref{eq:13} is entirely inertial, it is possible to employ the covariant formulation by initially establishing the affine connection without gravity \cite{DZ}.\\
Furthermore the mathematical expression for the electromagnetic energy-momentum tensor $\varepsilon_{i j}$ is given by,
$$
\varepsilon_{i j}=2\left(F_{i k} F_{j k}-\frac{1}{4} g_{i j} F_{k l} F^{k l}\right),
$$
In the above expression,
$$
F_{i j}=\mathcal{A}_{i, j}-\mathcal{A}_{j, i}.
$$

The electromagnetic field tensor can be written as,

\begin{eqnarray}
 F_{i j, k}+F_{k i, j}+F_{j k, i}=0, \\ \label{Er}
 \left(\sqrt{-g} F^{i j}\right)_{, j}=\frac{1}{2} \sqrt{-g} j^i.
\end{eqnarray}

The electromagnetic four potential is denoted by $\mathcal{A}_i$, whereas the four current density is represented by $j^i$. In the context of a static fluid arrangement and with consideration of spherical symmetry, the sole component of the four-current density that exhibits a nonzero value is denoted as $j^0$ is oriented along the radial direction $r$. Hence, with the exception of the radial component $F_{01}$ of the electric field, the static and spherically symmetric nature of the electric field implies the vanishing of all other components of the electromagnetic field tensor. If the condition $F_{01}=-F_{10}$, which implies antisymmetry, is satisfied, then Equation (\ref{Er}) is satisfied. The electric field equation can be obtained from equation (\ref{Er}) as :

$$
E(r)=\frac{1}{2 r^2} e^{\lambda(r)+\nu(r)} \int_0^r \sigma(r) e^{\lambda(r)} r^2 d r=\frac{q(r)}{r^2},
$$
In this context, the symbol $\sigma$ represents the charge density, while $q(r)$ represents the overall charge of the system.

\section{Modified Field Equation in $f(Q,T)$ gravity}\label{III}
For the current analysis, we are going to consider the static spherically symmetric metric configuration for describing the inner structure of the compact object,
\begin{eqnarray}\label{eq:metric}
    ds^2=-e^{\nu(r)} dt^2+e^{\lambda(r)} dr^2+r^2(d\theta^2+sin^{2}\theta \, d\phi^2),
\end{eqnarray}
which provides the distance formula $ds^2=g_{\mu\nu}dx^{\mu}dx^{\nu}$, where $x^{\nu}=(t,r,\theta,\phi)$ is the components of the four-dimensional space-time and $\nu(r)$ and $\lambda(r)$ represents the static metric potential along the time and radial co-ordinate respectively. Moreover, we model the dense matter by an isotropic perfect fluid whose components of energy-momentum tensor are given by $(-\rho,p,p,p)$
where $\rho$ is the density of the matter distribution, $p$ is the pressure of the fluid. Furthermore, by following the references \cite{1,2}, we have chosen the matter Lagrangian $L_{m}=p$. Consequently, the components of $\Phi_{\mu\nu}$ can be derived from the expression $\Phi_{\mu\nu}=g_{\mu\nu}p-2T_{\mu\nu}$. The effective terms in \eqref{field} can be written as,
\begin{gather*}
    \rho^{\text{eff}}=\rho+\alpha \Theta^{0}_{0}\, , \quad p_r^{\text{eff}}=p-\alpha \Theta^{1}_{1}\, , \quad p_t^{\text{eff}}=p-\alpha \Theta^{2}_{2},
\end{gather*}
and the corresponding anisotropy factor for the effective term is,
\begin{eqnarray}
    \Delta^{\text{eff}}=p_t^{\text{eff}}-p_r^{\text{eff}}=\alpha\left(\Theta^{1}_{1}-\Theta^{2}_{2}\right).
\end{eqnarray}



Now, the non-metricity scalar is defined as 
\begin{eqnarray}
    Q = - \frac{2 e^{-\lambda(r)}(r \nu'(r)+1)}{r^2}.
\end{eqnarray}
For the metric \eqref{eq:metric} the three non-vanishing independent field equations in $f(Q,T)$ gravity are given by,


\begin{eqnarray}
    \label{FE1}
\rho^{\mathrm{eff}}+\frac{n}{2}(3\rho-p)+\frac{q^2}{r^4} &=& -m\Big[e^{-\lambda}\big(\frac{\lambda^{\prime}}{r}-\frac{1}{r}\big)+\frac{1}{r^2}\Big],\\
\label{FE2}
p_r^{\mathrm{eff}}+\frac{n}{2}(3 p-\rho) -\frac{q^2}{r^4} &=& -m\Big[e^{-\lambda}\big(\frac{\nu^{\prime}}{r}+\frac{1}{r^2}\big)-\frac{1}{r^2}\Big],\\ 
 \label{FE3}
 p_t^{\mathrm{eff}}+\frac{n}{2}(3p-\rho) + \frac{q^2}{r^4} &=& -m \Big[e^{-\lambda}\Big\{\frac{\nu^{\prime \prime}}{2}+\big(\frac{\nu^{\prime}}{4}+\frac{1}{2 r}\big)\times \nonumber\\&&\hspace{-0.5cm} \left(\nu^{\prime}-\lambda^{\prime}\right)\Big\}\Big].
\end{eqnarray}

Our next target is to solve the above field equation and find an exact solution that perfectly describes the decoupled strange star model. 
In the above system of field equation, we have chosen the functional form of $f(Q,T)$ as, $f(Q,T)=m\,Q+n\,T$. 
This particular form of $f(Q, T)$ has been studied in various literature for reference \cite{Yix}. Additionally, under a specific transformation along the gravitational potential, we will employ the method known as gravitationally decoupling through the MGD technique~\cite{JO2}.
\begin{eqnarray}\label{trns1}
    \nu(r) &=& a(r)+\alpha \, g(r), \\\label{trns2}
    e^{-\lambda(r)} &=& b(r)+\alpha \,h(r).
\end{eqnarray}

Here, $\alpha$ is the decoupling constant. In this specific situation, the manipulation of the geometry involves decoupling certain functions in the $t-t$ and $r-r$ components. Specifically, we represent the geometric deformation along the $r-r$ component as $h(r)$ and the deformation along the $t-t$ component as $g(r)$. In this scenario, we are only interested in modifying the $r-r$ component while keeping the $t-t$ component unchanged. Therefore, we set $g(r)=0$. This choice reflects the primary objective of this manipulation, which is to change only one of the metric potentials. This approach is known as MGD. In this particular context, the mass function $\Tilde{m}(r)$ can be expressed as follows:

\begin{eqnarray}\label{mf}
    &&\tilde{m}(r)=\underbrace{ 4\pi \int_0^{r} \rho(\tilde{r}) \Tilde{r}^2 d\Tilde{r} +\frac{1}{2} \int_0^{r} \frac{q^2(\Tilde{r})}{\Tilde{r}^2}d\Tilde{r}+\frac{q^2 (\Tilde{r})}{2\Tilde{r}}}_{\Tilde{m}_{G R}(r)}\nonumber \\&&\hspace{0.5cm}+\underbrace{\int_0^{r}\frac{n}{4}\left(3\rho-p\right) \Tilde{r}^2 d\Tilde{r}}_{\Tilde{m}_{f Q T}(r)}+\underbrace{4\pi \alpha \int_0^{r} \Theta_0^{0}(\Tilde{r}) \Tilde{r}^2\,d\Tilde{r}}_{\Tilde{m}_{M G D}(r)} .
\end{eqnarray}

Eventually, by setting $n=0$ and $\alpha=0$, we can get the usual mass function for the charged compact object in the context of GR. From the Eq.~(\ref{mf}), one can see we have denoted the mass function as $\Tilde{m}_{fQT}(r)$ and $\Tilde{m}_{MGD}(r)$, which is coming from the mass contribution of $f(Q,T)$ and MGD technique respectively.

Now, plugging the transformation (\ref{trns1},\ref{trns2}) and using the corresponding functional form of $f(Q,T)$ mentioned previously, in Eqs.(\ref{FE1}-\ref{FE3}), we get two sets of equations. The first system is derived by setting $\alpha=0$, yields the subsequent standard field equations, 
\begin{eqnarray}\label{a1}
   \rho+\frac{n}{2}\left(3\rho-p\right) +\frac{q^2}{r^4} &=&  -\frac{m}{r^2}\left(1-b(r)\right)+\frac{m\,b^{\prime}(r)}{r},\\
   \label{a2}
   p+\frac{n}{2}\left(3p-\rho\right)-\frac{q^2}{r^4} &=& \frac{m}{r^2}\left(1-b(r)\right)-\frac{m\,b(r)a'(r)}{r},\\
   \label{a3}
   p+\frac{n}{2}(3 p-\rho)+\frac{q^2}{r^4} &=& -m\Big[\frac{b(r)a''(r)}{2}\nonumber +\Big(\frac{a'(r)}{4}+\frac{1}{2r}\Big)\\&&\big(b(r)a'(r)+b'(r)\big)\Big].
\end{eqnarray}
The above set of equations (\ref{a1}-\ref{a3}) is the solution of the following space-time metric :\\
\begin{eqnarray}
    ds^2=-e^{a(r)} dt^2+ b^{-1}(r) dr^2+r^2(d\theta^2+sin^2\theta d\phi^2).~~~~
\end{eqnarray}
Under these circumstances, the gravitational mass of the charged compact object can be written as,

\begin{eqnarray}
    m_0(r) &=& \underbrace{4\pi \int_0^r \rho(r) r^2 dr+\frac{1}{2}\int_0^r\frac{q^2}{r^2} dr+\frac{q^2}{2r}}_{m_{GR}(r)}\\&& \hspace{-0.3cm} + \underbrace{\frac{n}{4}\int_0^r (3\rho-p)r^2 dr}_{m_{fQT}(r)}.
\end{eqnarray}
The second set of equations, which includes the extra component for the decoupling section, results in

\begin{eqnarray}\label{b1}
    \Theta^0_{0} &=& m\left(\frac{h}{r^2}+\frac{h'}{r}\right),\\ \label{b2}
    \Theta_{1}^{1} &=& m \left(\frac{a'(r)h}{r}+\frac{h}{r^2}\right),\\ \label{b3}
    \Theta^{2}_{2} &=&  m \left[\frac{h a''(r)}{2}+
    \frac{a'(r)^2h}{4}+\frac{h a'(r) h'}{4}+\frac{h'}{2r}\right].
\end{eqnarray}




Additionally, From the Eqs.(\ref{a2},\ref{a3}) the electric charge $q^2$ for the strange star in isotropic case under $f(Q,T)$ gravity can be derived as,

\begin{eqnarray}\label{el}
    \frac{2q^2}{r^4} = -\frac{m}{4 r^2} \Big\{r \big(2 b r a''+a' (b r a'+r b'-2 b)+2 b'\big)\nonumber\\-4( b-1)\Big\}.~~~~
\end{eqnarray}

Now, from the Eqs. (\ref{a1}-\ref{a3}), we get the explicit form of isotropic pressure and density given by,
\begin{widetext}
    \begin{eqnarray}\label{pre}
    \rho &=&\frac{m \left(r \left(2 b (n+1) r a''+a' \left((n+1) r b'-2 (3 b n+b)\right)+b (n+1) r \left(a'\right)^2+2 (7 n+5) b'\right)+4 (b-1) (n+1)\right)}{8 (n+1) (2 n+1) r^2},\\ \label{den}
    p &=& \frac{m \left(r \left(-2 b (n+1) r a''-b' \left((n+1) r a'-2 n+2\right)-b a' \left((n+1) r a'+10 n+6\right)\right)-4 (b-1) (n+1)\right)}{8 (n+1) (2 n+1) r^2}.
\end{eqnarray}

\end{widetext}

\section{Matching condition for the astrophysical system}\label{IV}
The stellar distribution at the surface ($r=r_{\Sigma}$) of any stable spherically symmetric celestial object must be smooth and continuous between the exterior solutions ($r>r_{\Sigma}$) and the interior ($r<r_{\Sigma}$). Due to their matching, the space-time geometry becomes physically feasible. Here we are considering the Reissner Nordstrom exterior space-time for describing the exterior space-time of the star, which is given by,
\begin{eqnarray}\label{+}
    dS_{+}^2 &=& -\Big(1-\frac{2 {\mathcal{M}}}{r}+\frac{\mathbb{{Q}}^2}{r^2}\Big)dt^2+\Big(1-\frac{2 {\mathcal{M}}}{r}+\frac{\mathbb{Q}^2}{r^2}\Big)^{-1} dr^2 \nonumber\\&&+r^2(d\theta^2-sin^2{\theta} d\phi^2).
\end{eqnarray}
The Reissner Nordstrom metric, which represents the gravitational field outside a non-rotating, charged, spherically symmetric body of mass $\mathcal{M}$, is a static solution to the Einstein-Maxwell field equations. 
However, the following line element provides the most general interior metric that includes the geometric deformation:
\begin{eqnarray}\label{-}
    dS_{-}^2 &=& -e^{a(r)} dt^2+[b(r)+\alpha h(r)]^{-1} dr^2 \nonumber\\&&+r^2(d\theta^2+sin^2\theta d\phi^2).
\end{eqnarray}
Now, for the sake of stable configuration, the inner manifold $dS_{-}^2$ (\ref{-}) must be smoothly matched with the outer manifold $dS_{+}^2$ (\ref{+}) at the boundary $\Sigma$. The well-known continuity equation, which finally yields the first and second fundamental forms across the surface $\Sigma$, is the process of combining both geometries at this boundary. Concerning the first fundamental form, the inner geometry represented by the metric tensor $g_{\mu\nu}$ generated on the interface by $dS_{-}^2$ and $dS_{+}^2$ can be seen as follows:

\begin{eqnarray}
g_{tt}^{-}|_{r=r_{\Sigma}}=g_{tt}^{+}|_{r=r_{\Sigma}}\,, \\
g_{rr}^{-}|_{r=r_{\Sigma}}=g_{rr}^{+}|_{r=r_{\Sigma}}\,.
\end{eqnarray}

From Eq.~(\ref{-}) and Eq.~(\ref{+}) it takes the explicit form as, 

\begin{eqnarray}\label{C}
    b(r)+\alpha h(r) &=& \left(1-\frac{2 {\mathcal{M}}}{r}+\frac{\mathbb{Q}^2}{r^2}\right),\\ \label{c2}
    e^{a(r)} &=& \left(1-\frac{2 \mathcal{M}}{r}+\frac{\mathbb{Q}^2}{r^2}\right).
\end{eqnarray}

On the other hand, the second fundamental form takes the form as,

\begin{eqnarray}
    p_r^{\text{eff}}(r)|_{\Sigma}=\left[p(r)-\alpha \,\Theta_{r}^{r}(r)\right]_{\Sigma}=0.
\end{eqnarray}
The outer spacetime $r>r_{\Sigma}$, defined by the electric field, indicates that the compact object is not submerged in a vacuum spacetime as we deal with charge configuration. Moreover, the $\Theta$ sector might theoretically introduce changes to the matter composition and exterior geometry. Regarding the electric field contribution, the electric charge at the surface interface $r_{\Sigma}$ needs to be continuous to provide a well-defined matching between the inner and outer manifolds. Thus
\begin{eqnarray}\label{ch4}
    q(r)|_{r=r_\Sigma}= \mathbb{Q}.
\end{eqnarray}

In the above equation, $Q$ is the total electric charge contained in the outer space-time and $q(r)$ represents the amount of electric charge confined into the fluid sphere. 
Naturally, the concurrence of the electric charge magnitude at the junction surface $r_{\Sigma}$ also implies preserving the electric field continuity across $r_{\Sigma}$. Considering the aforementioned factors, the second fundamental form assumes its ultimate formulation as :
\begin{eqnarray}
p\left(r_{\Sigma}\right)-\alpha\left[\Theta_{1}^{1}\left(r_{\Sigma}\right)\right]^{-} = -\alpha\left[\Theta_{1}^{1}\left(r_{\Sigma}\right)\right]^{+},
\end{eqnarray}

Here, $p(r_{\Sigma})=p(r_{\Sigma})^{-}$.

 Now plugging the form of $\left(\Theta_{1}^{1}\right)^{-}$ and $\left(\Theta_{1}^{1}\right)^{+}$ by using the Eq.~(\ref{b1}) and Eq.~(\ref{c2}) into the above equation, we get the following expression,


\begin{eqnarray}\label{22}
p\left(r_{\Sigma}\right)- m\,\alpha  \Big[ h(r_{\Sigma})\big(\frac{a^{\prime}(r_{\Sigma})}{r_{\Sigma}}+\frac{1}{r_{\Sigma}^{2}}\big)\Big] &=& -m\, \alpha h^{*}(r_{\Sigma})\nonumber\\&& \hspace{-5cm}\Big[\frac{2}{r_{\Sigma}^{2}} \big(\frac{\mathcal{M}}{r_{\Sigma}}-\frac{\mathbb{Q}^{2}}{r_{\Sigma}^{2}}\big) \frac{1}{\big(1-\frac{2 \mathcal{M}}{r_{\Sigma}}+\frac{\mathbb{Q}^{2}}{r_{\Sigma}^{2}}\big)}+\frac{1}{r_{\Sigma}^{2}}\Big],
\end{eqnarray}

where 
\begin{eqnarray}
\left. a^{\prime}\left(r_{\Sigma}\right) \equiv \partial_{r} a^{-}\right|_{r=r_{\Sigma}}.
\end{eqnarray}
In the above expression $h^{*}\left(r_{\Sigma}\right)$ is represented by the deformation function in the context of the exterior Reissner-Nordstrom solution in response to the presence of an additional source denoted as $\Theta_{\mu \nu}$. For the Reissner-Nordstrom solution to give the exterior solution, it is necessary to satisfy the condition that $h^{*}(r_{\Sigma})=0$. Thus,
\begin{eqnarray}
    p(r_{\Sigma})- m\,\alpha  \Big[ h(r_{\Sigma})\big(\frac{a^{\prime}\left(r_{\Sigma}\right)}{r_{\Sigma}}+\frac{1}{r_{\Sigma}^{2}}\big)\Big]=0.
\end{eqnarray}

The above condition can be also written as
\begin{eqnarray}\label{jun}
    p\left(r_{\Sigma}\right)-\alpha\left(\Theta_{1}^{1}\left(r_{\Sigma}\right)\right)^{-}=0.
\end{eqnarray}

\section{Class I Condition and its solution in $f(Q,T)$ gravity via MGD approach}\label{V}

\subsection{Fundamental of the class condition}
The metric represents a 4-dimensional spherically symmetric spacetime that generally characterizes a Class II spacetime. This demonstrates the necessity of utilizing a 6-dimensional pseudo-Euclidean space for the purpose of embedding. Alternatively, it is possible to identify a viable parametrization that allows the entanglement of the space-time into a pseudo-Euclidean space with five dimensions. In the given scenario, the Class variable will have a value of $p=1$, referred to as embedding Class I. It is widely recognized that for any given $m$-dimensional spacetime $S_{m}$, there exists an isometric embedding of $S_{m}$ in a flat space of $m(m+1) / 2$ dimensions. However, in the case where the sum of $m$ and $n$ is the minimum order dimension of this flat space, the notation $V_{m}$ is referred to as an embedding class $n$ spacetime. Any spherically symmetric spacetime, whether static or non-static, must meet the following necessary and sufficient characteristics in order be considered as a Class I:

\begin{itemize}
  \item The following relation must be established for a system of symmetric values $b_{\mu \nu}$:
\end{itemize}

\begin{eqnarray}
    R_{\mu \nu \alpha \beta}=\epsilon\left(b_{\mu \alpha} b_{v \beta}-b_{\mu \beta} b_{\nu \alpha}\right), \text{(Gauss' equation)},
\end{eqnarray}

Where $\epsilon$ will take the value $+1$ or $-1$ whenever the normal to the manifold is space-like or time-like, respectively.
\begin{itemize}
  \item The symmetric tensor $b_{\mu \nu}$ need to meet the following differential equations :
\end{itemize}
\begin{eqnarray}
    \nabla_{\alpha} b_{\mu \nu}-\nabla_{\nu} b_{\mu \alpha}=0, \quad \text { (Codazzi's equation) }
\end{eqnarray}

Furthermore, the terms for the Riemannian components under the space-time metric in Schwarzschild's coordinates $(t, r, \theta, \phi \equiv$ $0,1,2,3$ ) can be calculated as :

\begin{eqnarray*}
    & R_{r t r t}=-e^{\nu}\left(\frac{\nu^{\prime \prime}}{2}-\frac{\lambda^{\prime} \nu^{\prime}}{4}+\frac{\nu^{\prime 2}}{4}\right) ; \quad R_{r \theta r \theta}=-\frac{r}{2} \lambda^{\prime} ; \\
& R_{\theta \phi \theta \phi}=-\frac{r^{2} \sin ^{2} \theta}{e^{\lambda}}\left(e^{\lambda}-1\right) ; \quad R_{r \phi \phi t}=0, \\
& R_{\phi t \phi t}=-\frac{r}{2} \nu^{\prime} e^{\nu-\lambda} \sin ^{2} \theta ; \quad R_{r \theta \theta t}=0 .
\end{eqnarray*}

Then, utilizing the above components in the Gauss equation, one can get :

\begin{eqnarray}
    b_{t r} b_{\phi \phi} = R_{r \phi t \phi}=0 ; \quad b_{t r} b_{\theta \theta}=R_{r \theta t \theta}=0, \,\,\\
    b_{t t} b_{\phi \phi}=R_{t \phi t \phi} ; \quad b_{t t} b_{\theta \theta} = R_{t \theta t \theta} ; \quad b_{r r} b_{\phi \phi}=R_{r \phi r \phi},\,\,\\
    b_{\theta \theta} b_{\phi \phi} = R_{\theta \phi \theta \phi} ; \quad b_{r r} b_{\theta \theta}=R_{r \theta r \theta} ; \quad b_{t t} b_{r r}=R_{t r t r}\,.\,\,
\end{eqnarray}

The following relation is obtained directly from the previous set of relations.

\begin{eqnarray}\label{Rie1}
    (b_{t t})^{2} =\frac{(R_{t \theta t \theta})^{2}}{R_{\theta \phi \theta \phi}} \sin ^{2} \theta, \quad(b_{r r})^{2}=\frac{(R_{r \theta r \theta})^{2}}{R_{\theta \phi \theta \phi}} \sin ^{2} \theta,~~~~ \\
     (b_{\theta \theta})^{2} = \frac{R_{\theta \phi \theta \phi}}{\sin ^{2} \theta}, \quad(b_{\phi \phi})^{2}=\sin ^{2} \theta R_{\theta \phi \theta \phi}.~~~~~
\end{eqnarray}

We obtain the following connection in Riemann components by substituting the above sets for the components of Eq.~(\ref{Rie1}).

\begin{eqnarray}
    R_{t \theta t \theta} R_{r \phi r \phi}=R_{t r t r} R_{\theta \phi \theta \phi}.
\end{eqnarray}

subject to $R_{\theta \phi \theta \phi} \neq 0$. It is important to point out that the above equations provide Codazzi's equation. One crucial aspect that we would want to highlight is that in the case of a non-static spherically symmetric spacetime, the symmetric tensor $b_{\mu \nu}$ will exhibit the following equation.

\begin{eqnarray}
    b_{t r} b_{\theta \theta}=R_{r \theta t \theta} \quad \text{and} \quad b_{t t} b_{r r}-\left(b_{t r}\right)^{2}=R_{t r t r}.
\end{eqnarray}

here $\left(b_{t r}\right)^{2}=\sin ^{2} \theta\left(R_{r \theta t \theta}\right)^{2} / R_{\theta \phi \theta \phi}$. In this secenario, the embedding Class I condition takes the following form :

\begin{eqnarray}
    R_{t \theta t \theta} R_{r \phi r \phi}=R_{t r t r} R_{\theta \phi \theta \phi}+R_{r \theta t \theta} R_{r \phi t \phi}.
\end{eqnarray}

However, in our particular scenario, the condition will be similar to that of the static spherically symmetric metric. The aforementioned requirement can be referred to as a necessary and sufficient condition for characterizing a spacetime as Class I. By incorporating the Riemann components into the given condition, we obtain the subsequent equation.

\begin{eqnarray}
    2 \frac{\nu^{\prime \prime}}{\nu^{\prime}}+\nu^{\prime}=\frac{\lambda^{\prime} e^{\lambda}}{e^{\lambda}-1}.
\end{eqnarray}

 where, $e^{\lambda} \neq 1$. The solution of the above differential equation requires spacetime to be a Class I. Now, integrating the above second-order differential equation yields the following relation between the gravitational potentials

\begin{eqnarray}
    e^{\lambda}=1+A \nu^{\prime 2} e^{\nu} \quad\text{or} \quad e^{\nu}=\left[B+C \int \sqrt{e^{\lambda}-1} d r\right]^{2}.~~~~~~\label{eq62a}
\end{eqnarray}

 where $A, B$ and $C$ are the integration constants. 
The above derived Karmarkar condition is dependent on the Riemannian components i.e it is dependent on the spherically symmetric metric, which is independent of theory of gravity.  Furthermore, Karmarkar condition gives a direct relation between two metric potential without solving the equations of motion. So, if one function is known then other metric function could be obtained through using the above relation~\ref{eq62a}.
It should be highlighted that the aforementioned methodology has been extensively applied in the study of compact structures representing actual celestial bodies \cite{em1,em2}, as well as in research related to dark matter \cite{em3} and wormhole solutions \cite{em4,em5}.

\section{Embedding CLASS I SOLUTION IN $f (Q, T)$ GRAVITY WITH MGD}\label{VI}
In this section, we investigate a class of charge compact stars solution using the embedding class I condition. It is worth noting that the selected seed spacetime represents a hybrid, with its temporal component derived from the well-known Adler metric \cite{em6} and its radial component from the Finch-Skea metric \cite{em7}. Additionally, this seed Class I spacetime was previously obtained within the framework of charged anisotropic fluid spheres in the context of General Relativity (GR) \cite{em8}. Therefore, we are writting the following class I metric, 
 \begin{eqnarray}
    dS^2=-(B+C r^2)^2 dt^2+(1+A r^2) dr^2\\ &&\hspace{-3.5cm}+r^2(d\theta^2+sin^2\theta d\phi^2),\nonumber
\end{eqnarray}

here $b(r)=(1+Ar^2)^{-1}$ and $a(r)=2 \, \text{ln}[B+Cr^2]$. The reason for choosing such a particular seed space-time solution is, it satisfies some mathematical and physical necessary conditions. For example, the time component of the space-time metric satisfies some necessary condition like $e^{a(r)}|_{r=0}=\text{finite and constant}$, $a'(r)|_{r=0}=0$ and $a''(r)|_{r=0}>0$ which imply that $a(r)$ must be finite and regular throughout the interior region.  It also gives the minimum value at the center of the sphere and increases towards the radius of the fluid sphere.

Now, by implementing the above seed solution into the Eqs.(\ref{pre}) and (\ref{den}) we get density and pressure profile as:

\begin{eqnarray}\label{pr}
    \rho = -\frac{m}{\mathcal{K}}\Big\{(A^2 (n+1) r^2 (B+C r^2)\nonumber\\&&\hspace{-4cm}+2 A (B (4 n+3)+C (7 n+4) r^2 \Big\},\\ \label{de}
    p = \frac{m}{\mathcal{K}}\Big\{(A^2 (n+1) r^2 (B+C r^2)\nonumber\\&&\hspace{-4cm}+2 A (B-C (5 n+2) r^2)-4 C (3 n+2))\Big\},
\end{eqnarray}
where $\mathcal{K}=2 (n+1) (2 n+1) \left(A r^2+1\right)^2 \left(B+C r^2\right)$.
On the other hand, the electric field for isotropic case in $f(Q,T)$ gravity can be derived from Eq.~(\ref{el}) is given by,

\begin{eqnarray}\label{ch}
   E=\frac{q^2}{r^4}=-\frac{A m r^2 \left(A \left(B+C r^2\right)-2 C\right)}{2 \left(A r^2+1\right)^2 \left(B+C r^2\right)} .
\end{eqnarray}
Since we have specified the seed solution for the $f(Q,T)$-system, then we will move on to the second system known as $\theta$-sector. In the previous studies, the authors have widely used the mimic-constraint approach to solve the $\theta$-sector in different modified gravity theories along with the GR. The mimic approach was developed by Ovalle~\cite{JO2} in basis of physical ground so that the model should satisfy all the realistic conditions. Furthermore, the $f(Q,T)$ field equation can be solved by using the equation of state between the $\Theta$-sector~\cite{Sharif1}. On the other hand, the $\Theta$-sector can be miminc with another well-known dark matter density profile~\cite{Maurya11}. But in this article, we use the mimic-constraint approaches, $\rho=\Theta^0_0$ and $p=\Theta^1_1$, to solve our system, which is discussed in the next sections:   \\\\ 

\section{MGD solution by mimicking of $\Theta$ sector}\label{VII}
\subsection{Model I: Mimicking Pressure Constraint}
It is worth mentioning that the junction condition of the exterior Reissner-Nordstrom space-time is compatible $p_{r_{\Sigma}} \sim\left(\Theta_{1}^{1}\left(r_{\Sigma}\right)\right)^{-}$ which can be seen from the Eq.~($\ref{jun}$). Therefore we choose
$\Theta_{1}^{1}(r)=p(r)$.
In this case, we obtain the deformation function $h(r)$ from the Eq.~(\ref{b2}) as,

\begin{eqnarray}\label{hr}
    h(r) &=& \frac{r^2}{2 (n+1) (2 n+1) \left(A r^2+1\right)^2 \left(B+5 C r^2\right)} \Big[\big(A^2 (n+1)  \nonumber\\&& r^2(B+C r^2)+2 A \left(B-C (5 n+2) r^2\right)-4 C (3 n+2)\big)\Big],~~~~~~~
\end{eqnarray}


Hence, the gravitational metric potential takes the form :

 \begin{eqnarray}
   e^{-\lambda(r)} &=& \frac{1}{1+A r^{2}}+\alpha h(r).
\end{eqnarray}

In this scenario, the effective term for density and pressure quantity could be written as,

\begin{eqnarray}
    \rho^{\text{eff}}(r) &=& \rho(r)+\alpha \Theta_0^0\, ,\\
    p_r^{\text{eff}}(r) &=& p(r)-\alpha \Theta_1^{1} = (1-\alpha) p(r),\\
    p_t^{\text{eff}}(r) &=& p(r)-\alpha \Theta_2^{2}.
\end{eqnarray}
In the above expression $\rho(r)$
and $p(r)$ is given in equations (\ref{pr}) and (\ref{de}). Apart from this, the solution for the rest of the $\Theta$ sector is given in Appendix-I.\\

Now, the continuity of the first fundamental form is given by,

\begin{widetext}
\begin{eqnarray}\label{f1}
1-\frac{2 \mathcal{M}}{r_{\Sigma}}+\frac{\mathbb{Q}^{2}}{r_{\Sigma}^{2}}  &=& \left(B+C r_{\Sigma}^{2}\right)^{2}, \\ \label{f2}
1-\frac{2 \mathcal{M}}{r_{\Sigma}}+\frac{\mathbb{Q}^{2}}{r_{\Sigma}^{2}}  &=& \frac{1}{1+A r_{\Sigma}^{2}}+\alpha \frac{r_{\Sigma}^2 \left(A^2 (n+1) r_{\Sigma}^2 \left(B+C r_{\Sigma}^2\right)+2 A \left(B-C (5 n+2) r_{\Sigma}^2\right)-4 C (3 n+2)\right)}{2 (n+1) (2 n+1) \left(A r_{\Sigma}^2+1\right)^2 \left(B+5 C r_{\Sigma}^2\right)}.
\end{eqnarray}
\end{widetext}

Moreover, the continuity of the second fundamental form gives us 
\begin{eqnarray}\label{f3}
    (1-\alpha)p(r_{\Sigma})=0\, \implies \,p(r_{\Sigma})=0.
\end{eqnarray}

Now, applying the continuity of the first fundamental forms (\ref{f1}) and (\ref{f2}) and the second fundamental form (\ref{f3}), we get the values of the constants as,

\begin{eqnarray}\label{B}
B = \frac{-A^2 (n+1) r_{\Sigma}^4+2 A (5 n+2) r_{\Sigma}^2+12 n+8}{2 \sqrt{\big(A r_{\Sigma}^2+1\big) \big(n \big(5 A r_{\Sigma}^2+6\big)+3 A r_{\Sigma}^2+4\big)^2}},~~~~~~
\end{eqnarray}

\begin{eqnarray}
\label{C2}
C = \frac{A \big(A (n+1) r_{\Sigma}^2+2\big)}{2 \sqrt{\big(A r_{\Sigma}^2+1\big) \big(n \big(5 A r_{\Sigma}^2+6\big)+3 A r_{\Sigma}^2+4\big)^2}}.~~~~~
\end{eqnarray}

\subsection{Solution II: Mimic density constraint :}

In this scenario, we consider the mimic constraints for density as $\Theta_0^0(r)=\rho(r)$. After that by using the Eq.~(\ref{b1}) and Eq.~(\ref{de}) we get the following differential form in $h(r)$:

\begin{flalign}
\label{deform2}
    h^{\prime}(r)+\frac{h}{r}&=-\frac{\splitfrac{m r \big(A^2 (n+1) r^2 \big(B+C r^2\big)+4 C n}{+2 A \big(B (4 n+3)+ C (7 n+4) r^2\big)\big)}}{2 (n+1) (2 n+1) \big(A r^2+1\big)^2 \big(B+C r^2\big)}
\end{flalign}

After solving the above Eq.~(\ref{deform2}) we get the following expression of deformation function:

\begin{eqnarray}\label{def}
    h(r) &=& \frac{-m A r^2 (A B (7 n+5)-C (9 n+7))}{4 (n+1) (2 n+1) \left(A r^2+1\right) (A B-C)}.
\end{eqnarray}

In the above integration, the integrating constant is set to 0 to avoid singularity issues in the center of the strange star. Apart from that, in Eq.~(\ref{def}), to avoid the singularity at the origin, we have expanded the Taylor series of $tan^{(-1)}(r)$ up to the linear order of $r$. So, the gravitational metric potential for $r-r$ component is given by,
\begin{eqnarray}
    e^{-\lambda}(r)= \frac{1}{1+Ar^2}+\alpha \frac{-m A r^2 (A B (7 n+5)-C (9 n+7))}{4 (n+1) (2 n+1) \left(A r^2+1\right) (A B-C)}.\nonumber
\end{eqnarray}

Moreover, the form of effective density and pressure are given by,
\begin{eqnarray}
    \rho^{\text{eff}}(r) &=& \rho(r)+\alpha \Theta_0^0 =(1+\alpha) \rho(r),\\
    p_r^{\text{eff}} &=& p(r)-\alpha \Theta^1_{1},\\
    p_t^{\text{eff}} &=& p(r)-\alpha \Theta^2_{2}.~~~~~~~~~
\end{eqnarray}

Now, from Eq (.~\ref{b2}) we get the expression of $\Theta_1^1$ as,

\begin{eqnarray}
    \Theta_1^1=-\frac{A m^2 \left(B+5 C r^2\right) (A B (7 n+5)-C (9 n+7))}{4 \left(2 n^2+3 n+1\right) \left(A r^2+1\right) (A B-C) \left(B+C r^2\right)},\nonumber
\end{eqnarray}

After that, applying the same technique, we get the value of the constants B and C. Due to the long-expression, we have given the solution for the $\Theta$ sector in Appendix II.


\section{Physical Analysis}\label{VIII}
In this section, we have discussed the nature of certain physical quantities to ensure the viability of the decoupled strange star model. We have examined such physical features among the cases $f(Q,T)$, and $f(Q,T)$+MGD by modifying the value of the model parameter. We have seen a satisfactory outcome in each case, and we obtain excellent behavior of those physical parameters in $f(Q,T)$+MGD as compared to the $f(Q,T)$ model.

\subsection{Charge and deformation function}

\begin{figure*}[htbp]
    \includegraphics[width=8cm, height=5cm]{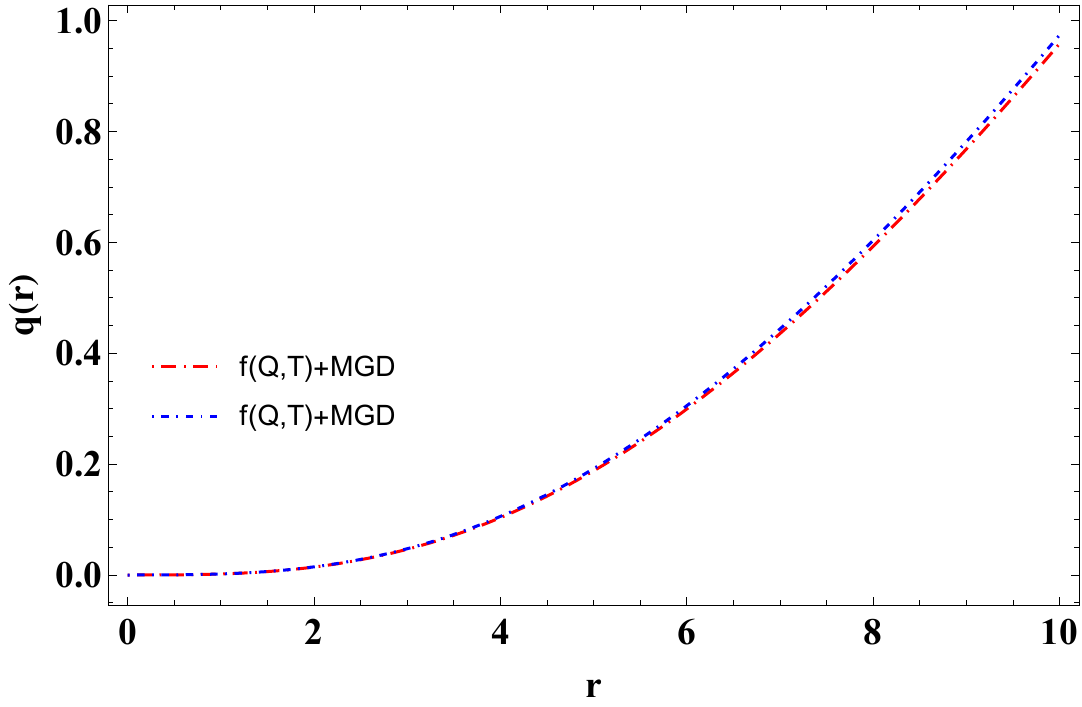}\,\,\,\,\,\,
      \includegraphics[width=8cm, height=5cm]{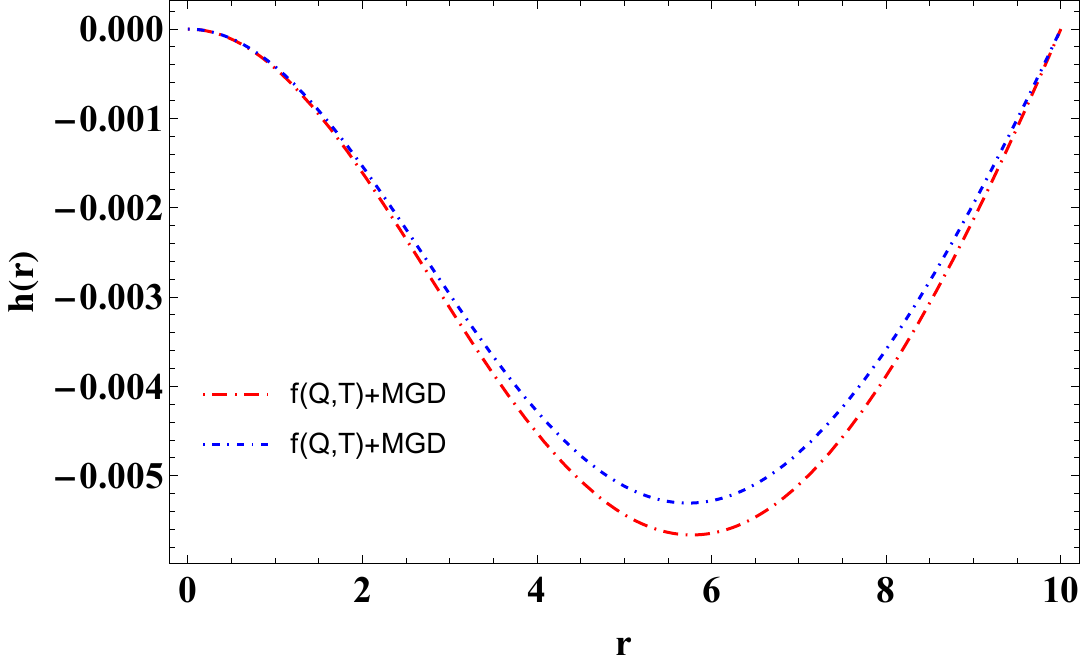}
       \caption{Graphical analysis of $q(r)$ and $h(r)$ are shown w.r.t $'r'$ for model-I, where $m=-0.1, A=0.01 $. For red and blue color lines $n=0.6, 0.8$ respectively. \label{hqr}}
\end{figure*}

\begin{figure*}[htbp]
    \includegraphics[width=8cm, height=5cm]{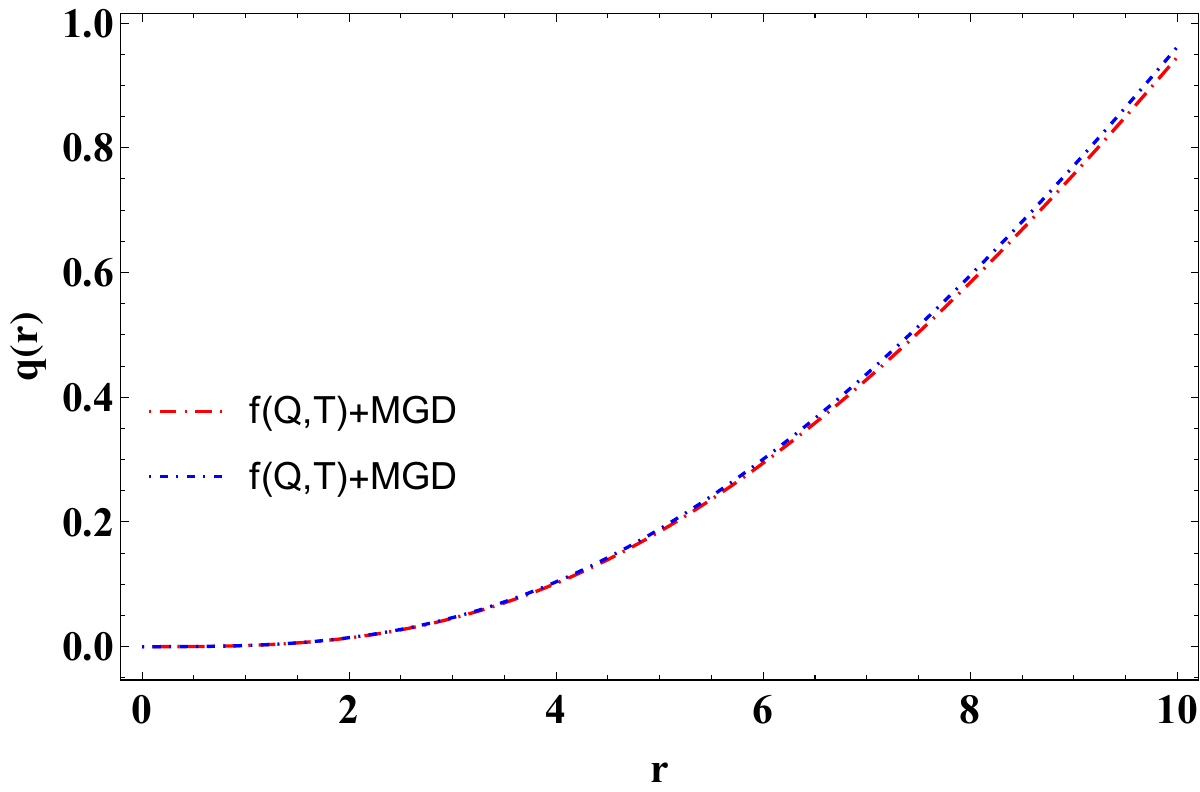}\,\,\,\,\,\,
      \includegraphics[width=8cm, height=5cm]{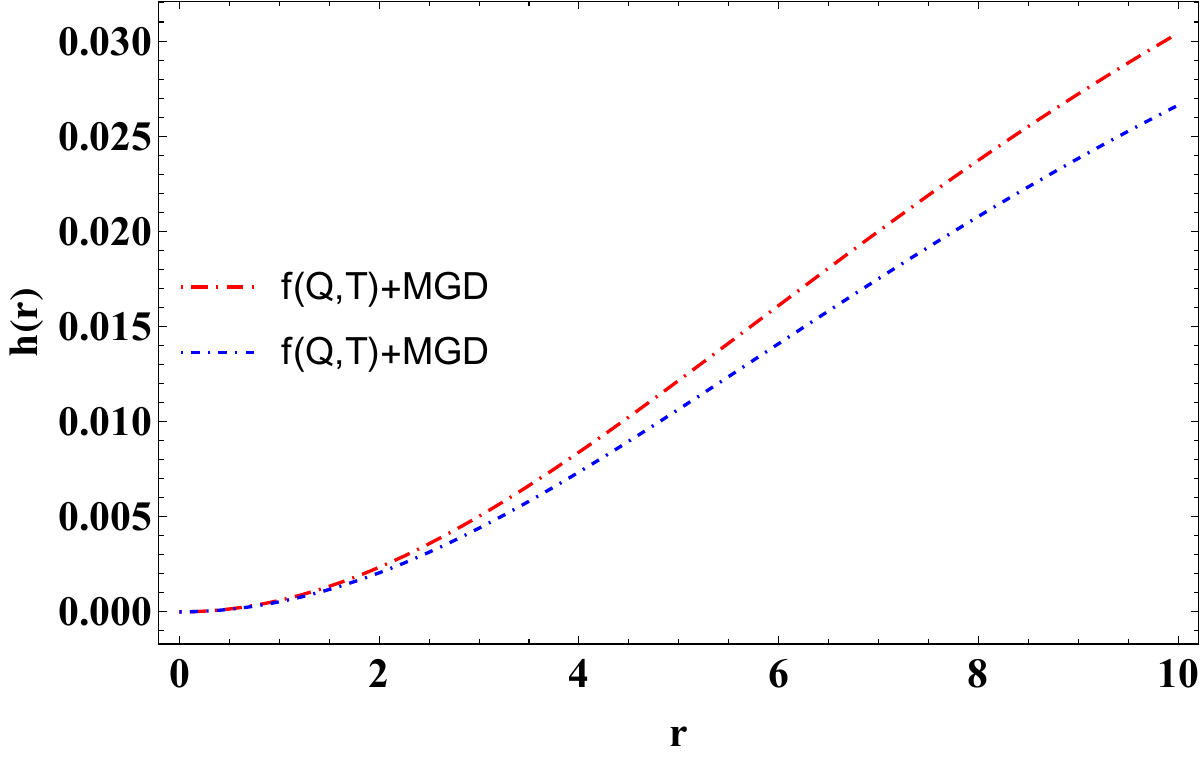}
       \caption{Graphical analysis of  $q(r)$ and $h(r)$ are shown w.r.t $'r'$ for model-II where $m=-0.1,  A=0.01, \alpha=-0.2$. For red and blue color lines, $n=0.6, 0.8$, respectively. \label{hqr2}}
\end{figure*}

The expression for the charge is given in  Eq.~(\ref{ch}), and the deformation function is provided for the two models in Eq.~(\ref{hr}) and (\ref{def}), respectively. We have shown the graphical analysis of the charge and deformation function in Figs.(\ref{hqr}) and (\ref{hqr2}) for the two models, respectively. There is no role of $\alpha$ in the charge and deformation functions for the solution-I. It should be noted that in solution-I, the function $h(r)$ exhibits a vanishing behavior at the boundary. That's why $h(r)$ does not affect the star's total mass for the solution-I. In this case, the total mass would be the same as the mass of the seed $f(Q,T)$ model. But, as evidenced by the solution-II depicted in Fig.~\ref{hqr2}, it is observed that the deformation function $h(r)$ exhibits a quick increase towards the star's boundary, hence influencing the total mass of the stellar configuration. In both cases, the charge function exhibits a monotonically increasing behavior. When the model parameter increases, the charge also increases towards the border, reaching a value of zero in the center.


\subsection{Nature of effective density, effective pressure, and anisotropic factor}
\begin{figure*}[htbp]
    \includegraphics[width=8cm, height=5cm]{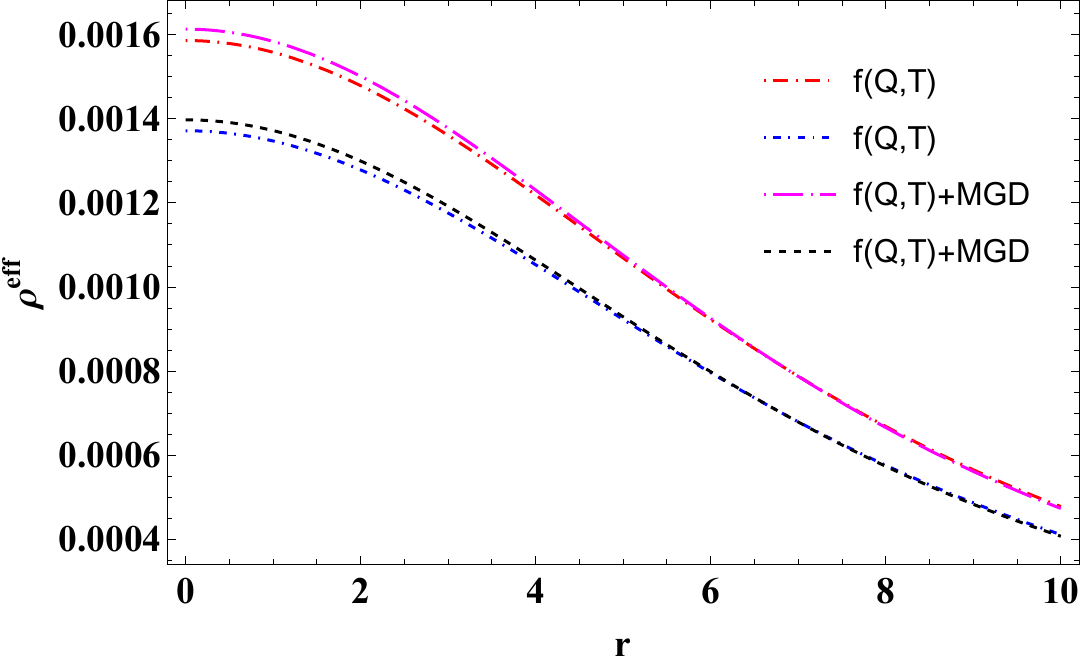}\,\,\,\,\,\,\,
      \includegraphics[width=8cm, height=5cm]{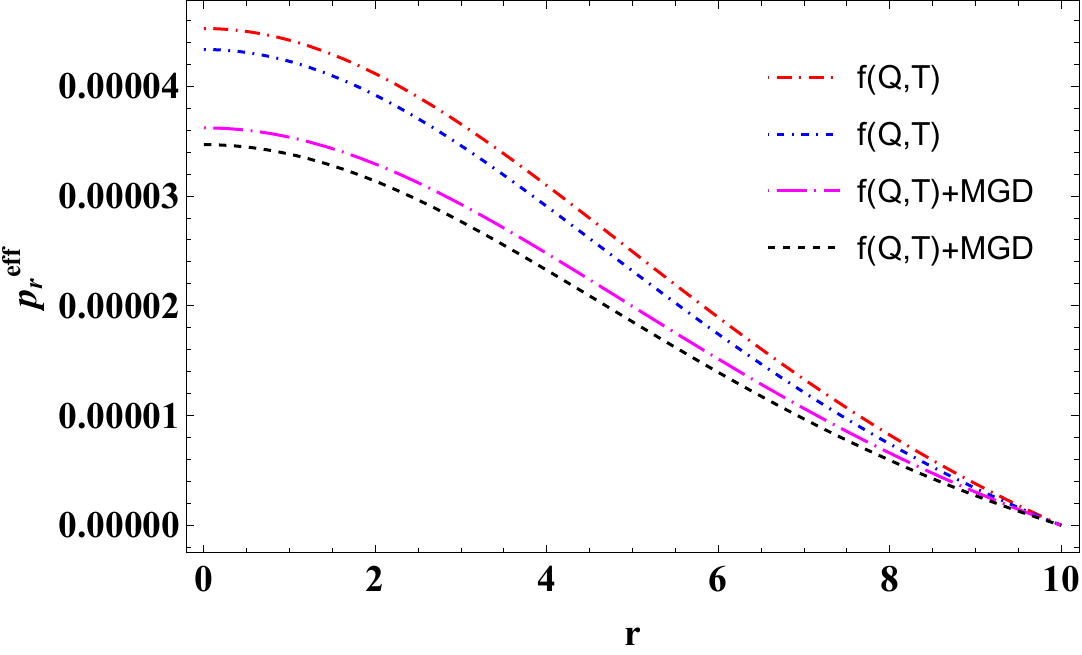}
    \includegraphics[width=8cm, height=5cm]{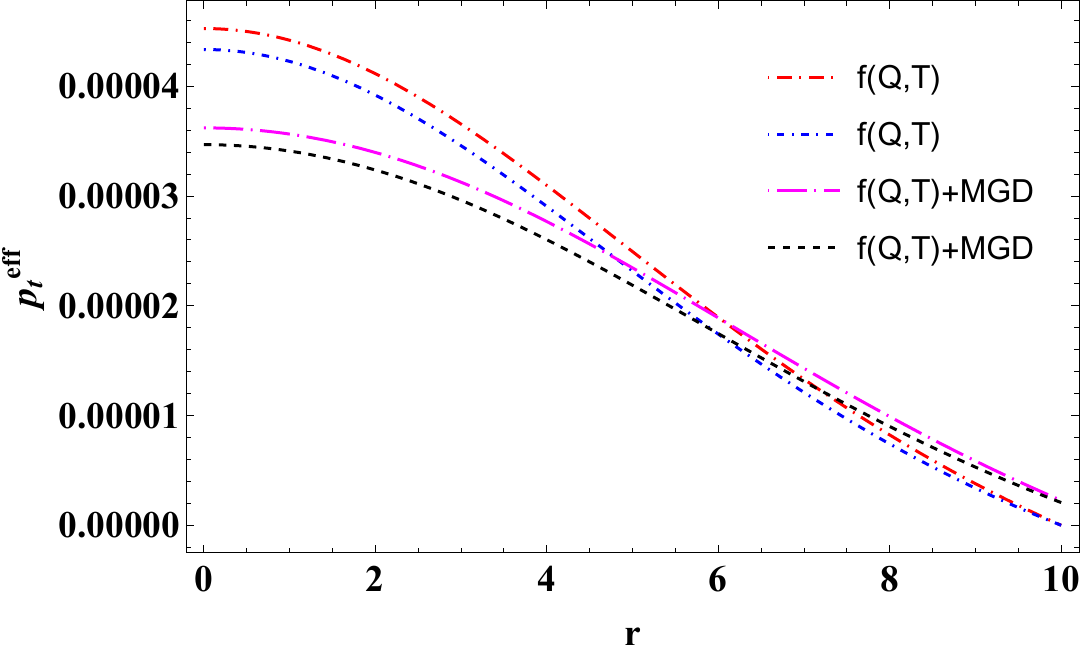}\,\,\,\,\,\,\,
      \includegraphics[width=8cm, height=5cm]{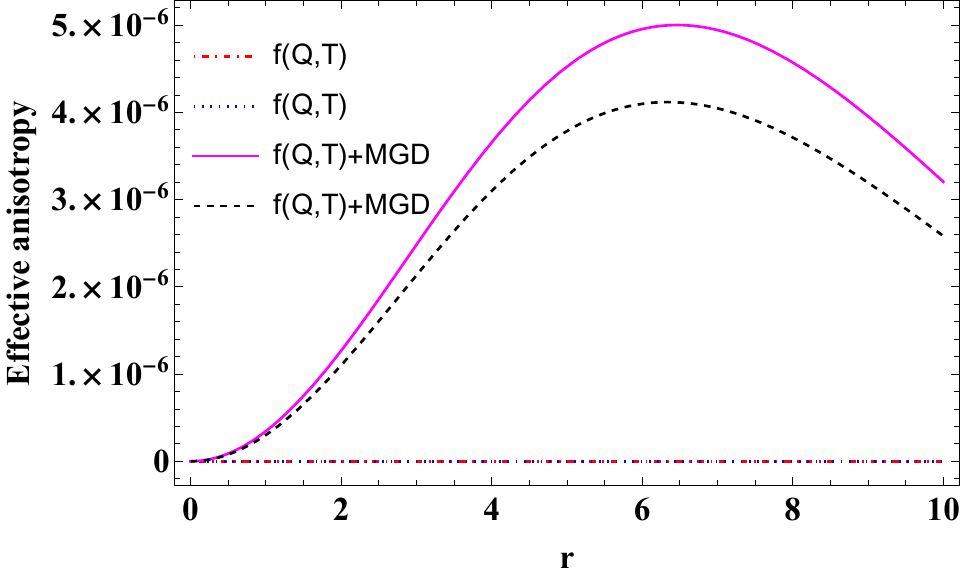}
       \caption{Variation of effective density, effective radial pressure, effective tangential pressure, and effective anisotropy w.r.t 'r' for model-I. Here $\alpha=0.2,A=0.01$. For red, blue, magenta and black color lines, $n=0.6, 0.8, 0.6,0.8$, respectively. \label{pt}}
\end{figure*}

\begin{figure*}[htbp]
    \includegraphics[width=8cm, height=5cm]{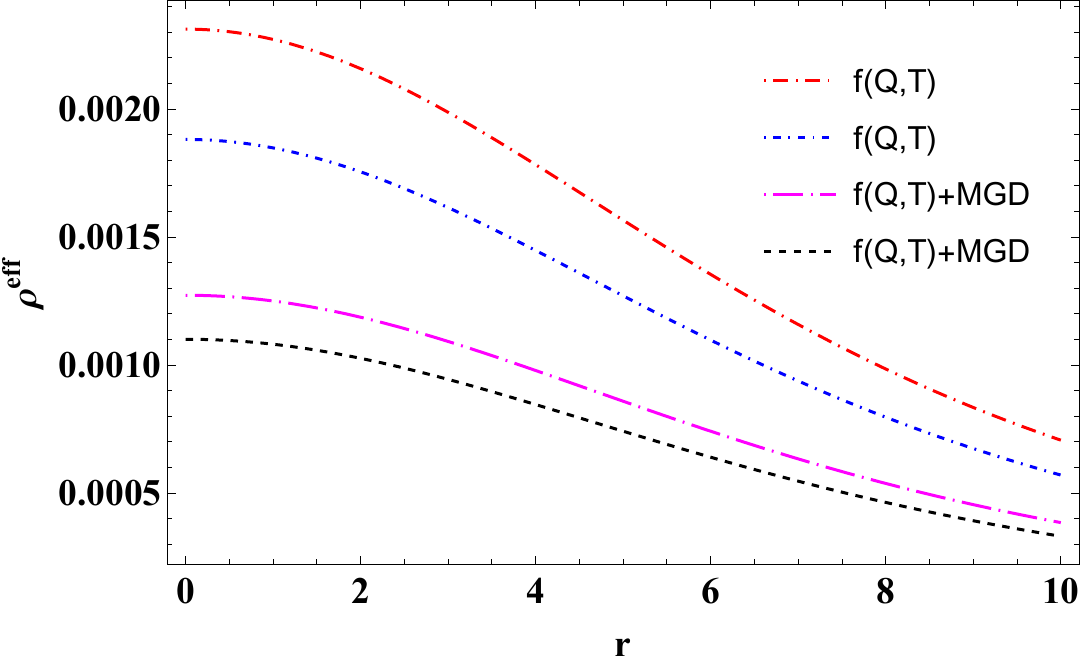}\,\,\,\,\,\,\,
      \includegraphics[width=8cm, height=5cm]{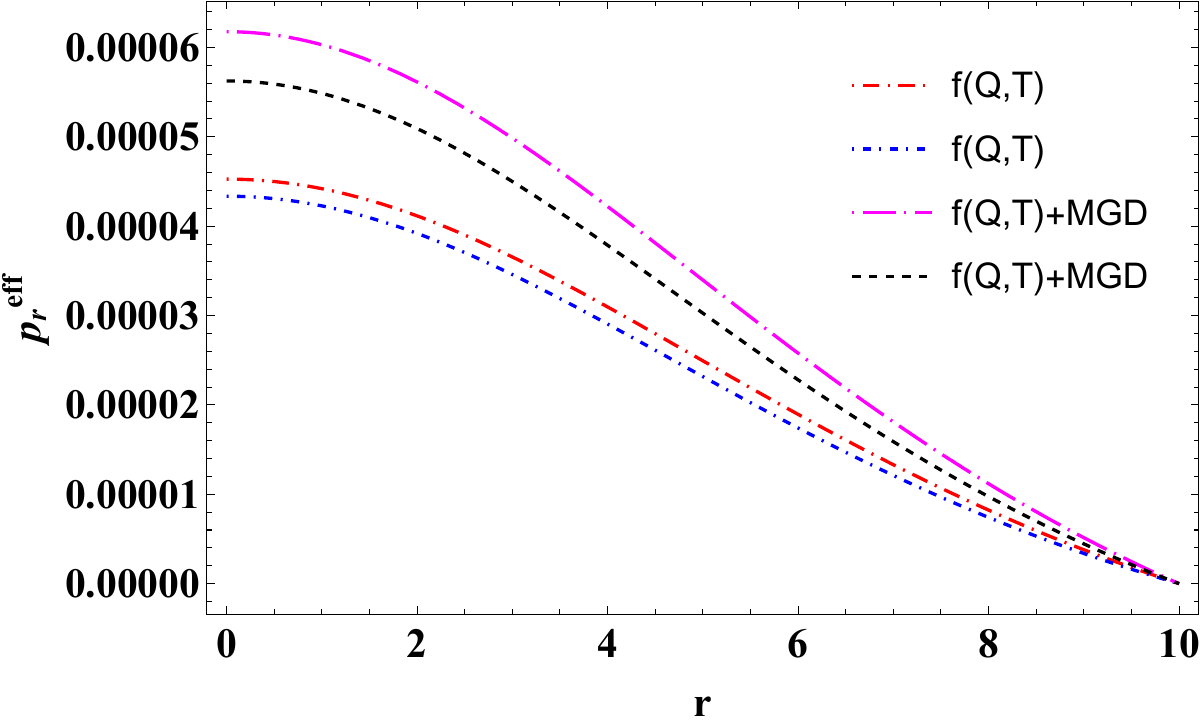}
    \includegraphics[width=8cm, height=5cm]{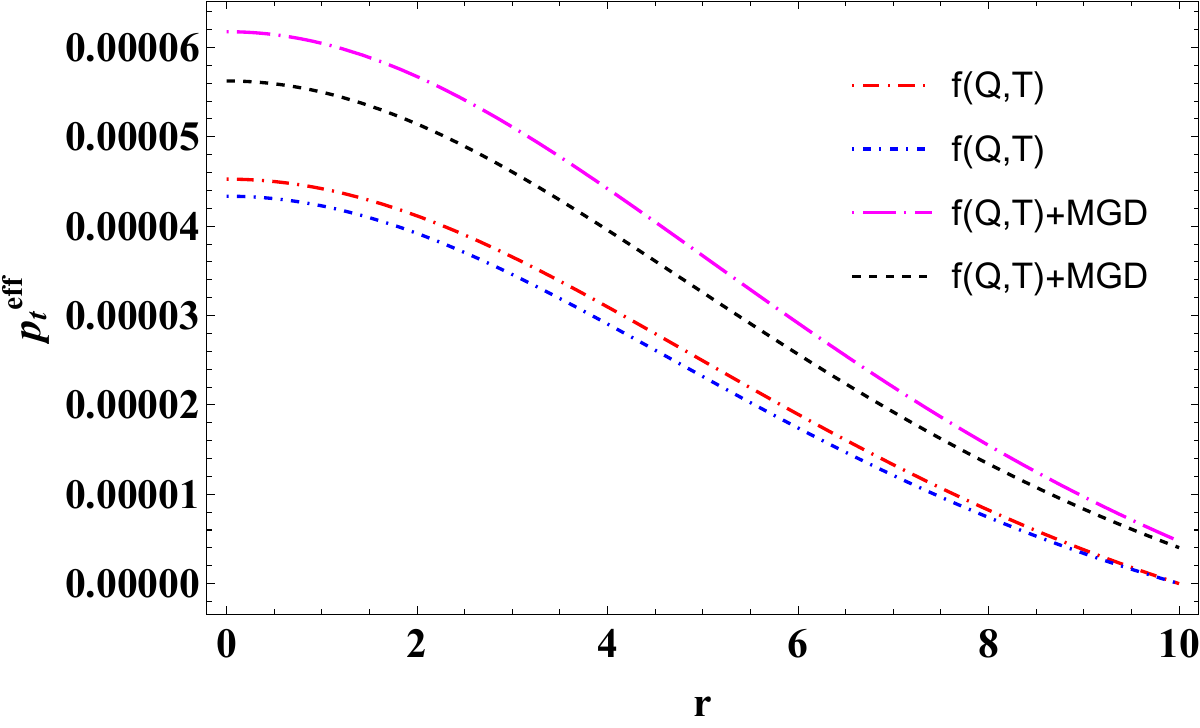}\,\,\,\,\,\,\,
      \includegraphics[width=8cm, height=5cm]{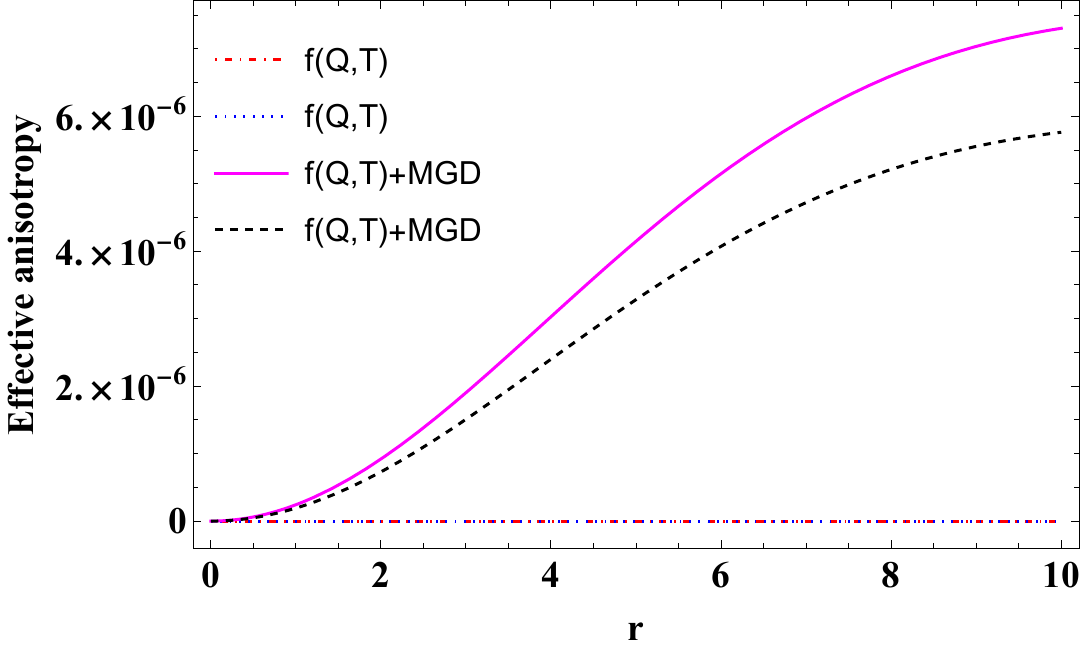}
       \caption{variation of effective density, effective radial pressure, effective tangential pressure and effective anisotropy. Here, $\alpha=-0.2, A=0.01$ for model II. For red, blue, magenta, and black lines, $n=0.6,0.8,0.6,0.8$ respectively.\label{pt2}}
\end{figure*}

This subsection focuses on the examination and analysis of the behavior of the three most significant aspects of the model, namely matter density, radial pressure, and tangential pressure. Furthermore, we investigate the role of the anisotropy factor $\Delta$ within the star's fluid. It is widely recognized that the main physical features of compact objects representing stellar interiors should not exhibit any physical or mathematical singularities. The matter density and pressure within a configuration should have the greatest values at its center while also displaying a monotonically decreasing trend as the radial coordinate approaches the surface. Some compact objects like quark stars, neutron stars, and white dwarfs need to be explained by these unique properties. Furthermore, one supplementary constituent holds equal significance in examining compact structures, providing a more precise depiction of the dynamics shown by celestial entities. For example, the material composition of the fluid sphere may have anisotropies. In the present context, anisotropy indicates that the pressures exerted in the radial and tangential directions are unequal, denoted as $p_r \neq p_t$. Hence, the anisotropy factor is defined as the difference between $p_t$ and $p_r$, denoted as $\Delta$.  It can be seen from Figs.(\ref{pt}) and (\ref{pt2}) that all the physical quantities remain finite and positive through the stellar radius. The effective pressure and effective density give the highest value at the center of the star, and after that, they are monotonically decreasing towards the boundary region. It is noted from Fig.(\ref{pt}) and (\ref{pt2}) that the effective density and effective pressure meet all the energy conditions. From the table (\ref{table1}), one can look at the central density, surface density, and central pressure of the stellar configuration.

\begin{figure*}[htbp]
    \includegraphics[width=8cm, height=5cm]{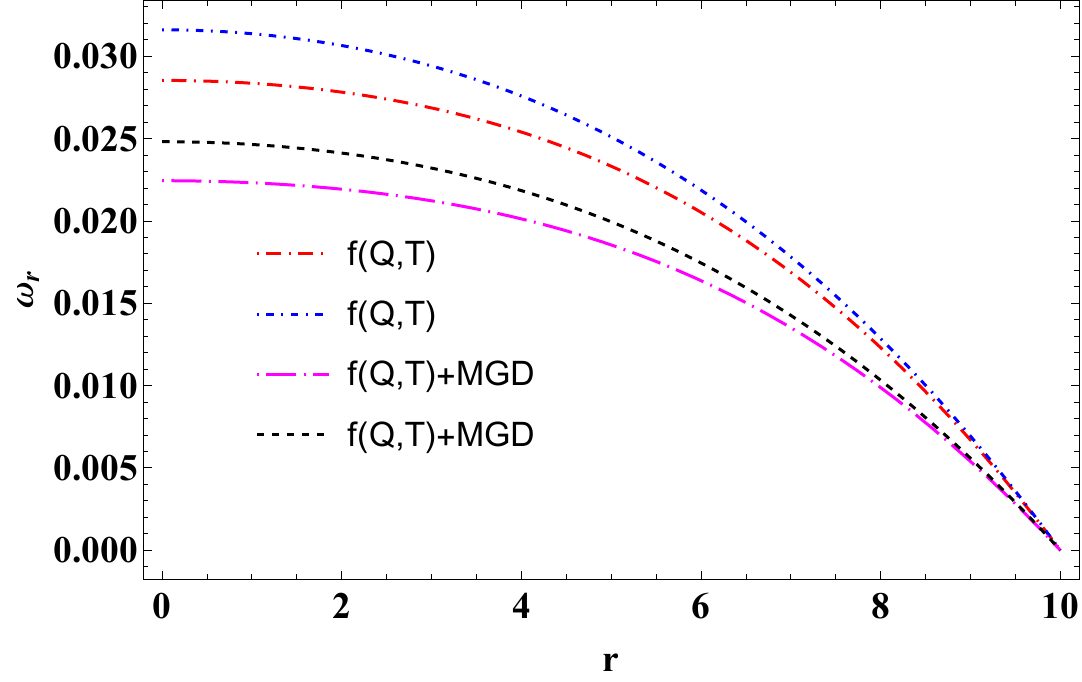}\,\,\,\,\,\,
      \includegraphics[width=8cm, height=5cm]{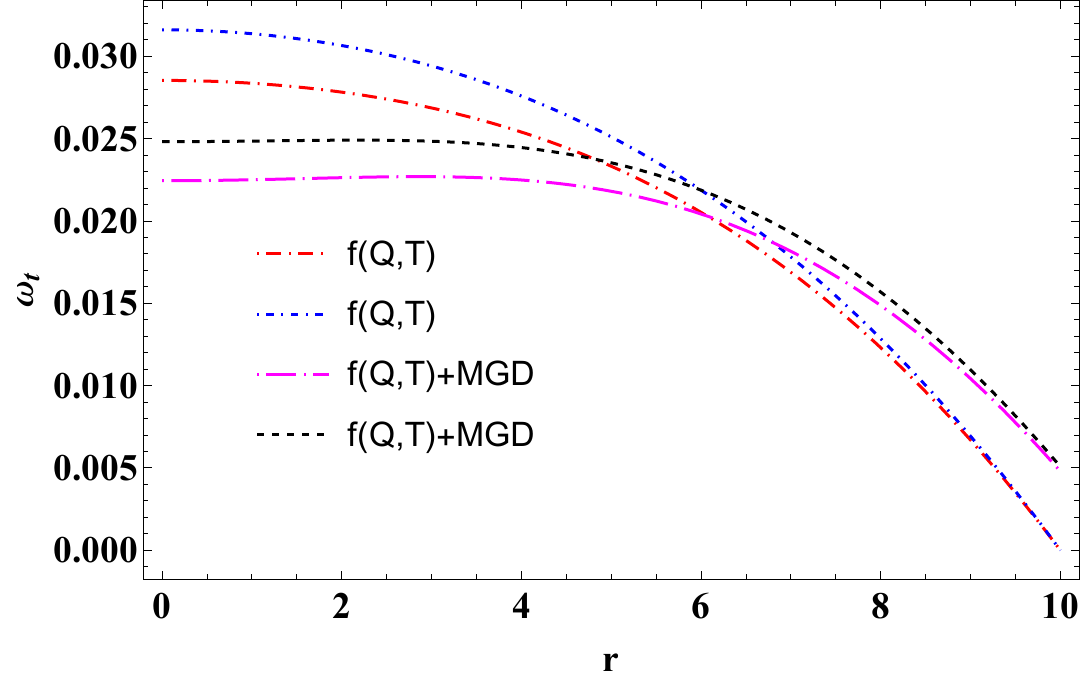}
       \caption{variations of $\omega_r$ and $\omega_t$ w.r.t $'r'$ for model-I. Here $\alpha=0.2,  A=0.01$ and for red, blue, magenta, and black lines, $n=0.6,0.8,0.6,0.8$ respectively.\label{eos}}
\end{figure*}

\begin{figure*}[htbp]
    \includegraphics[width=8cm, height=5cm]{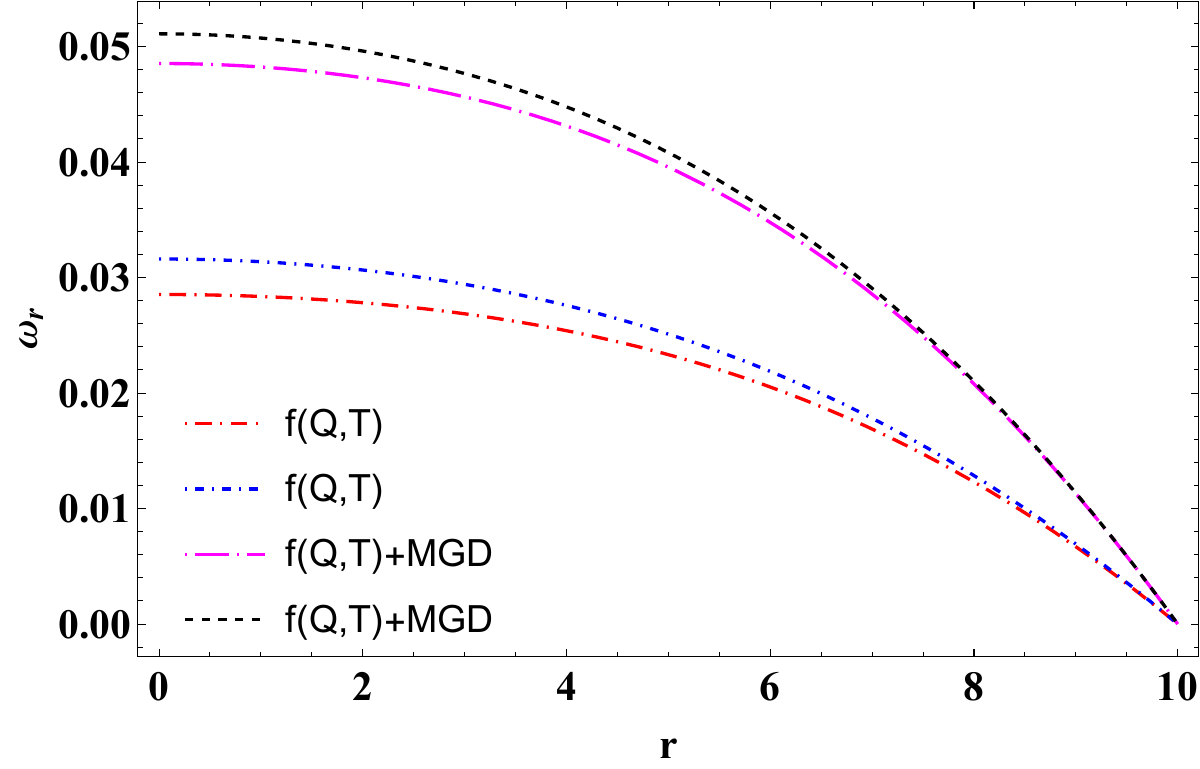}\,\,\,\,\,\,
      \includegraphics[width=8cm, height=5cm]{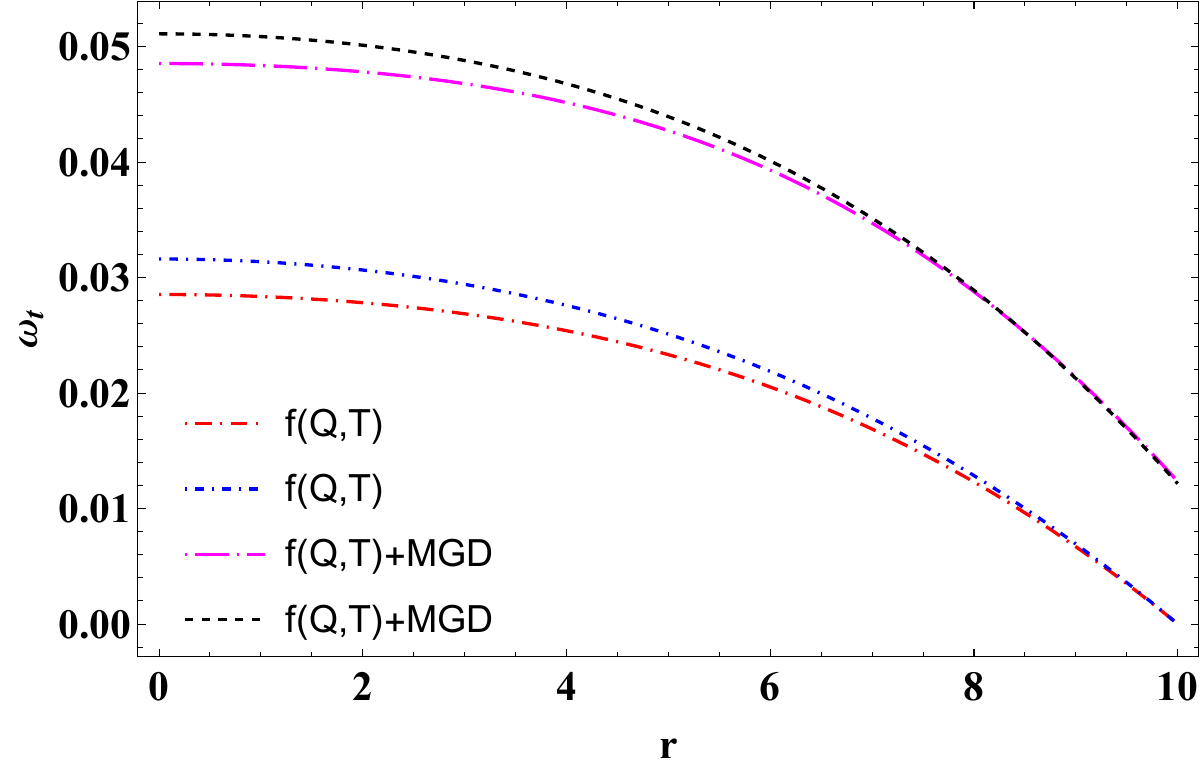}
       \caption{Variations of $\omega_r$ and $\omega_t$ w.r.t 'r' for model-II . Here $\alpha=-0.2, A=0.01$ and for red, blue, magenta, and black lines, $n=0.6,0.8,0.6,0.8$ .\label{eos2}}
\end{figure*}

\subsection{Equation of state parameter}

Another important way to characterize the relationship between matter density and pressure is by finding the equation of state parameters.  
The formula for the equation of state in our current model is given by:

\begin{eqnarray}
    \omega_r=\frac{p_r^{\text{eff}}}{\rho^{\text{eff}}},\, \quad
    \omega_t=\frac{p_t^{\text{eff}}}{\rho^{\text{eff}}}.
\end{eqnarray}
For any physical model, the $\omega_r=p_r/\rho$ and $\omega_t=p_t/\rho$ must be lies between 0 and 1 which is called a Zel'dovich condition. This means that model should satisfy the dominant energy condition i.e. $\rho-p_i\ge 0$ and velocity of sound can not be faster than the velocity of light.
The graphical analysis for the equation of state for both of the models is given in Figs.~\ref{eos} and \ref{eos2}, respectively. Our results indicate that both physical quantities give the highest value around the star's center and diminish towards the outside of the star. Furthermore, they are inside the bounds of the radiation era, i.e., $0\leq \omega_r, \, \omega_t \leq 1$.

\subsection{ Surface redshift}\label{IX}

\begin{figure*}[htbp]
    \includegraphics[width=8cm, height=5cm]{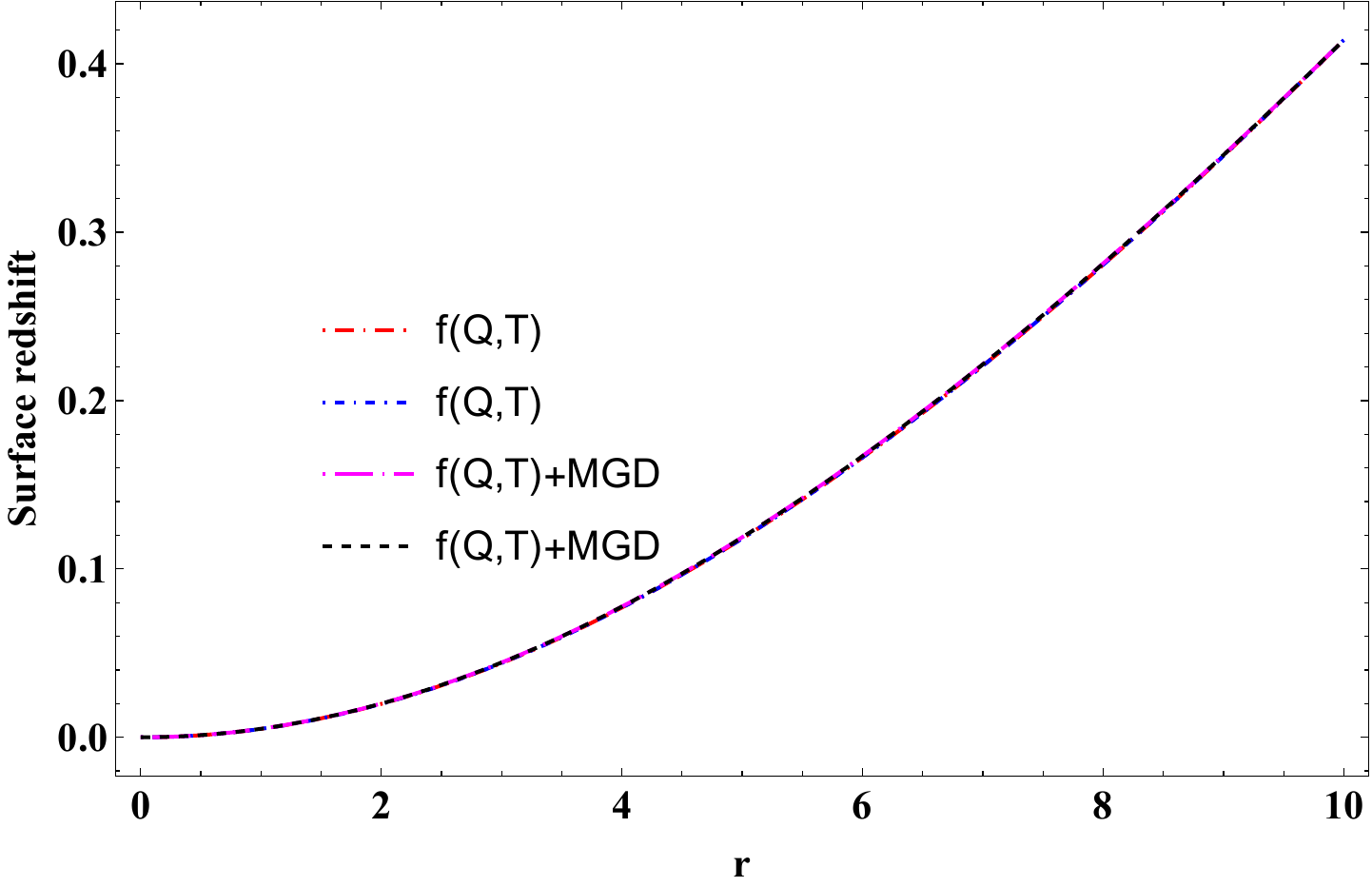}~~~~~~
     \includegraphics[width=8cm, height=5cm]{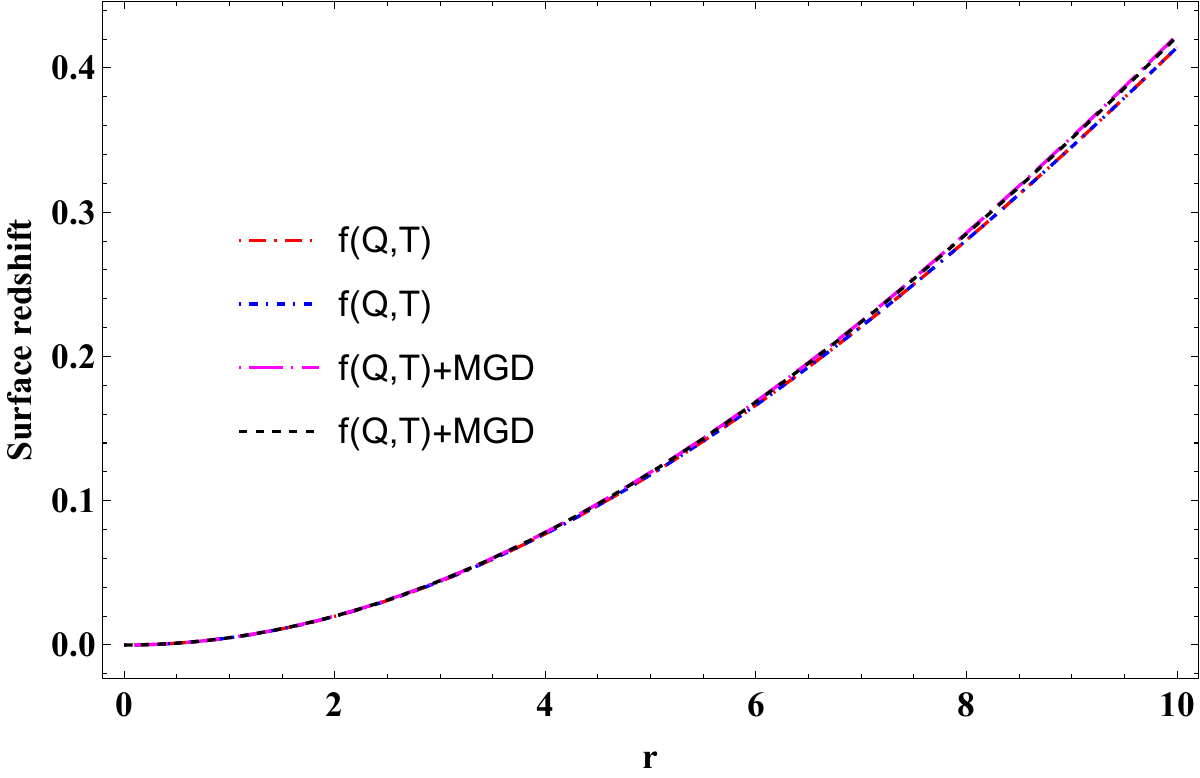}
    \caption{Variation of  surface redshift w.r.t 'r'. Here, $m =-0.1, \alpha=0.2$ for Model-I and, $m =-0.1, \alpha=-0.2$ for Model-II. Red, blue, magenta, black lines represent for $n=0.6,0.8,0.6,0.8$ respectively. \label{mass2}}
\end{figure*}

The surface redshift $\mathcal{Z}_{\mathrm{s}}$ is an important observable quantity that connects the mass and radius of a compact star, and it is determined by the formula $\mathcal{Z}_{\mathrm{s}}=\left(g_{tt}\right)^{-1/2}-1$.
Buchdahl \cite{Buchdahl1} suggested that the upper limit for the value of surface redshift should not exceed 2 for an isotropic, constant, perfect fluid distribution. However, according to Ivanov \cite{ivanov}, it might reach 3.84 for anisotropic fluid dispersion. From Fig.~\ref{mass2}, it is observable that the surface redshift is monotonically growing, but it does not exceed the refereed upper limit.\\

\begin{figure*}[htbp]
    \includegraphics[width=8cm, height=5cm]{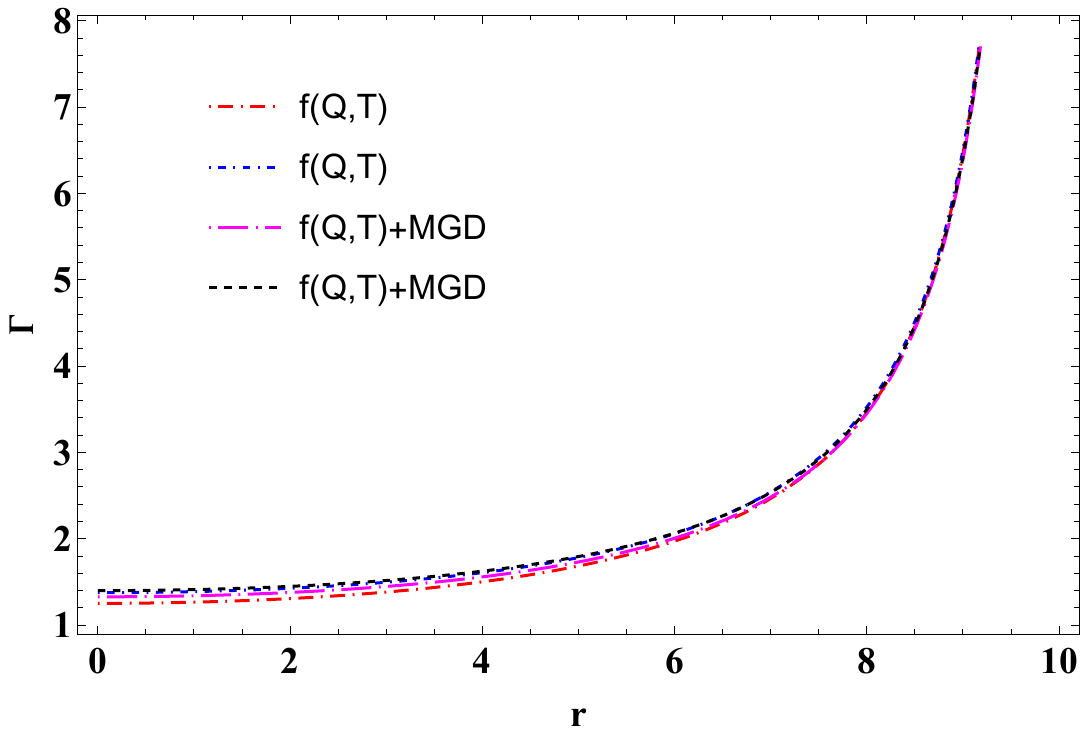}~~~~~
    \includegraphics[width=8cm, height=5cm]{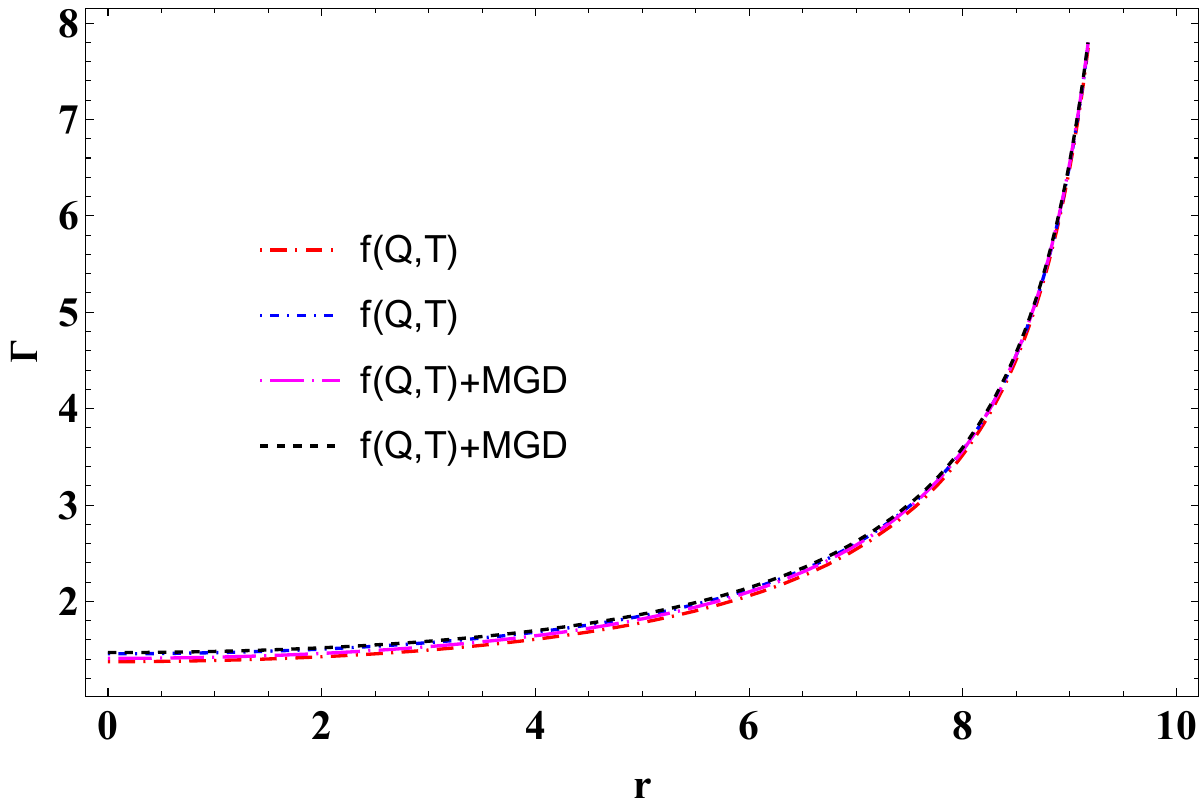}
    \caption{A comparative analysis of $\Gamma$ w.r.t 'r' for model-I (left) ($\alpha=0.2$) and model-II ($\alpha=-0.2$) (right). Here, $A=0.01$ and for red, blue, magenta, and black lines, $n=0.6,0.8,0.6,0.8$ respectively.\label{adb2}}
\end{figure*}

\begin{figure*}[htbp]
    \includegraphics[width=8cm, height=5cm]{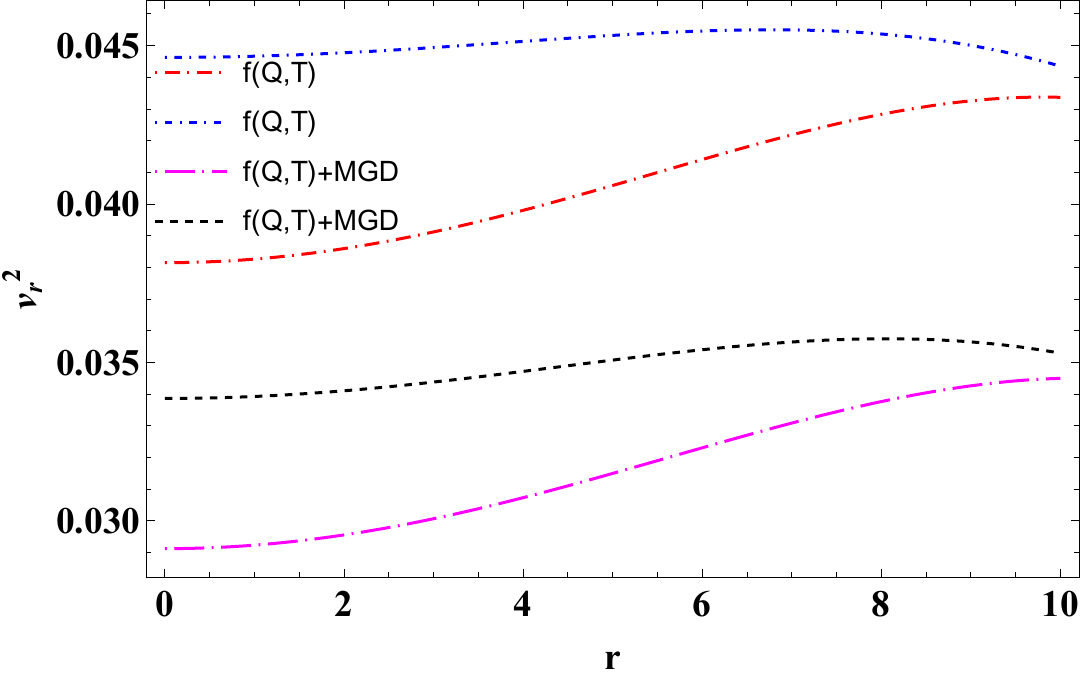}\,\,\,\,\,\,\,
      \includegraphics[width=8cm, height=5cm]{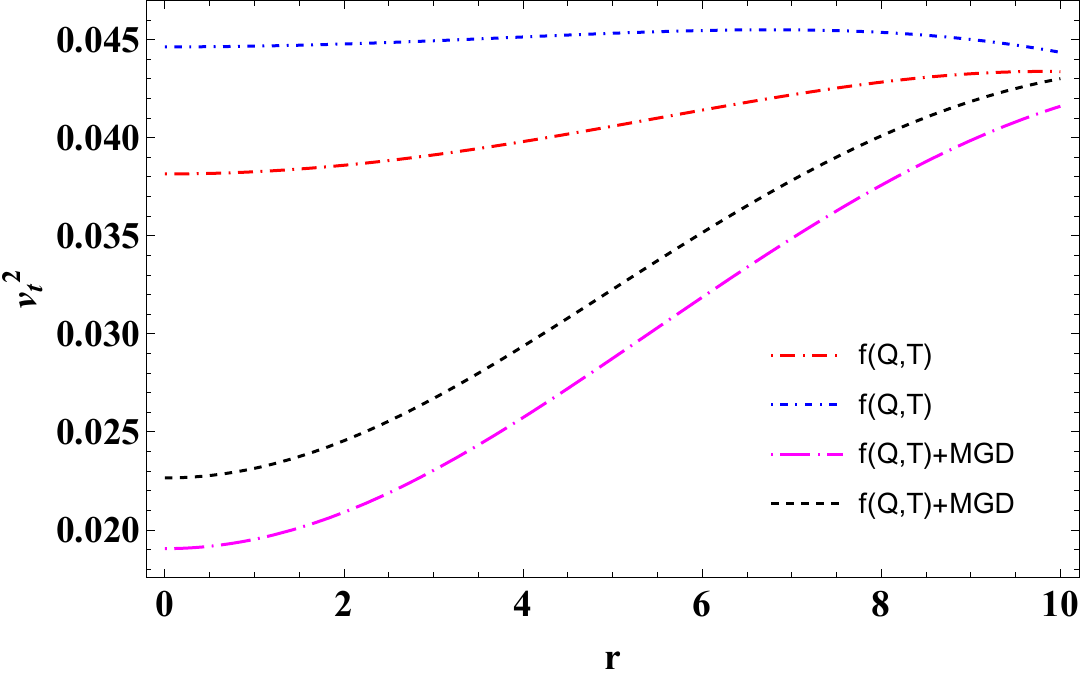}
       \caption{Graphical analysis of  $v_r^2$, $v_t^2$ w.r.t 'r' for model-I. Here $\alpha=0.2,  A=0.01$ and for red, blue, magenta, and black lines, $n=0.6,0.8,0.6,0.8$ respectively. \label{sound}}
\end{figure*}

\begin{figure*}[htbp]
    \includegraphics[width=8cm, height=5cm]{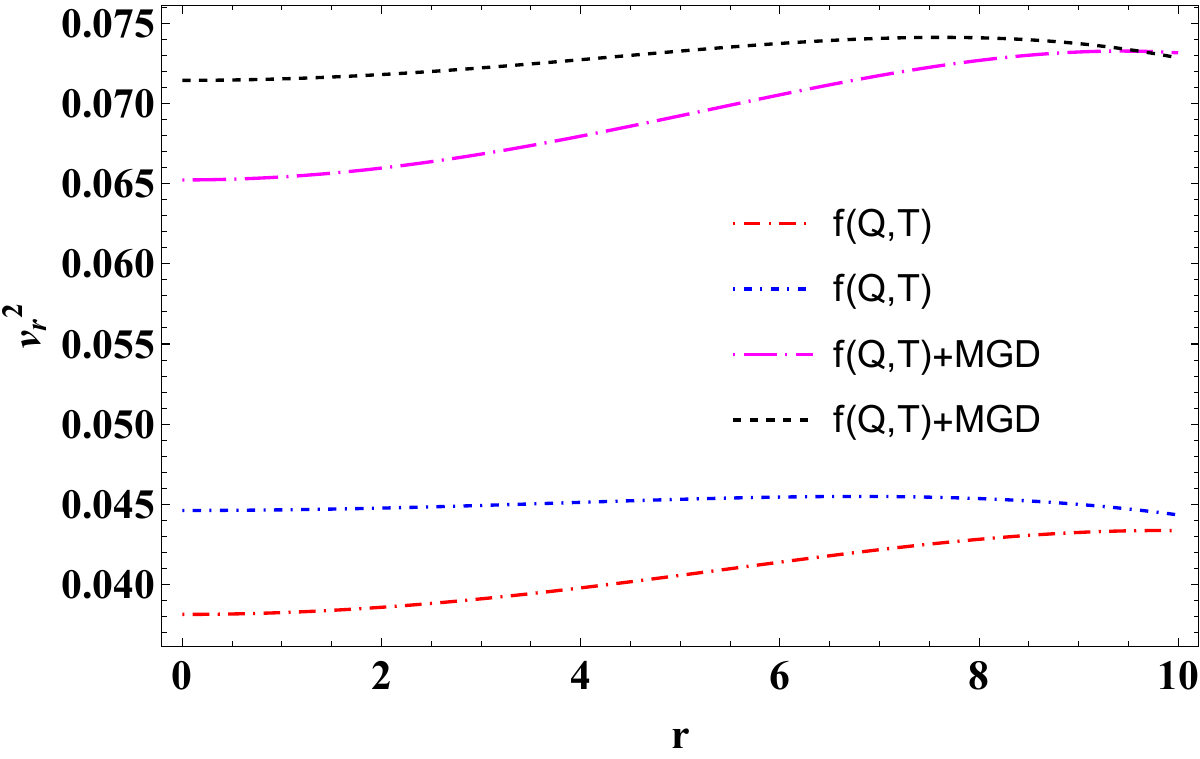}\,\,\,\,\,\,\,
      \includegraphics[width=8cm, height=5cm]{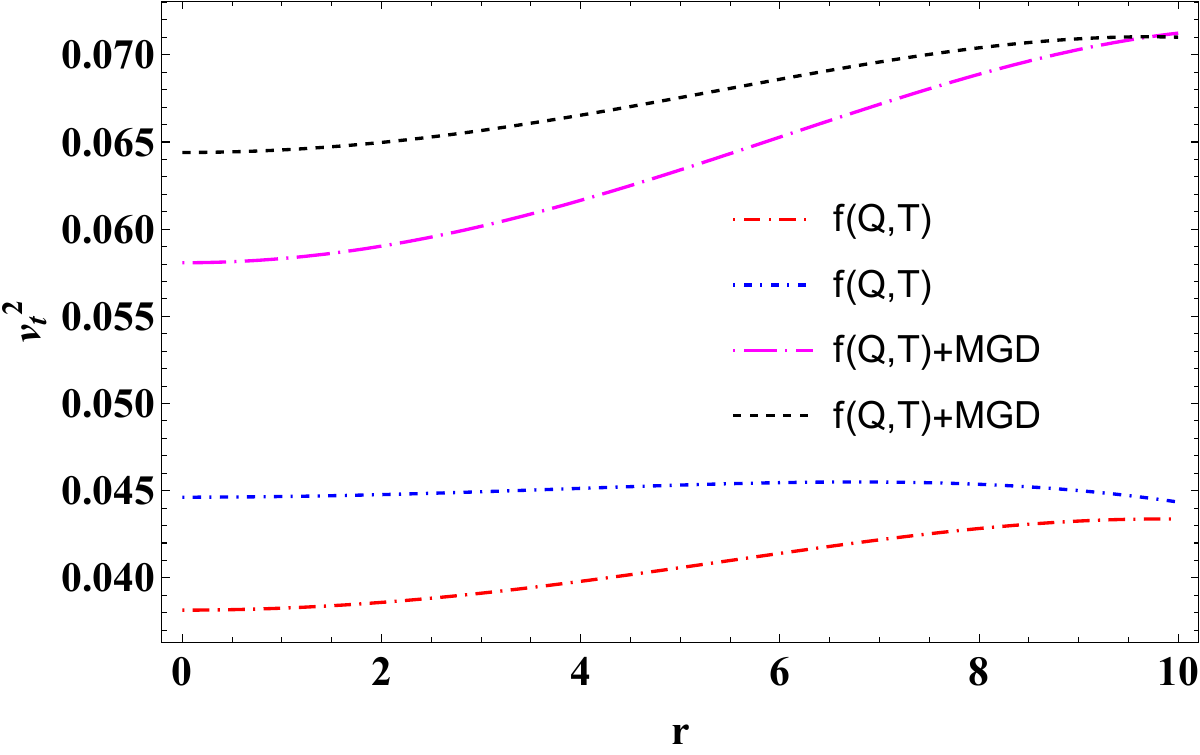}
       \caption{Graphical analysis of  $v_r^2$, $v_t^2$ w.r.t 'r' for model-II. Here $\alpha=-0.2,  A=0.01$ and for red, blue, magenta, and black lines, $n=0.6,0.8,0.6,0.8$ respectively.  \label{sound2}}
\end{figure*}

\section{Stability Analysis}\label{XI}
Now, in this section, we will evaluate the stability of our existing model by applying (i) the adiabatic index and (ii) the causality requirement. For examining the stability criteria, we have compared those physical properties among the cases $f(Q,T)$, and $f(Q,T)$+MGD.

\subsection{Adiabatic Index}
In this paragraph, we will conduct an analysis of a critical and significant ratio between the specific temperatures provided by $\Gamma$, to investigate the region of stability within the decoupled strange star model. The notion of the adiabatic index for an isotropic fluid sphere was introduced by Chan et al. \cite{chan}. However, Chandrasekhar \cite{chz} was among the early researchers to investigate the application of the adiabatic index in analyzing the stability region of spherical stars. The formula for the adiabatic index is given by,

\begin{eqnarray}
    \Gamma = \frac{\rho^{\text{eff}}+p_r^{\text{eff}}}{p_r^{\text{eff}}} \frac{dp_r^{\text{eff}}}{d\rho^{\text{eff}}}.
\end{eqnarray}

According to the study by Heintzmann and Hillebrandt \cite{hh}, the stellar object is considered stable when the value of the aforementioned expressions exceeds 4/3.

We have plotted the graph of the adiabatic index in Fig.~\ref{adb2} for both models. It satisfies the above-mentioned stability criterion for our constructed model in $f(Q,T)$ gravity with the MGD technique. 
One can carefully observe that, for a particular n, the value of the adiabatic index is higher than the above-mentioned limit.  In addition, our developed model demonstrates improved performance when the model parameter $n$ increases and the value of $\alpha$ decreases. For finding the range of $\alpha$, we have fixed $n=0.6$ for $f(Q,T)$+MGD case and varied the value of $\alpha$. The lower limits of the $\Gamma$ are $1.33$ and $1.42$ for $\alpha=0.2,0.1$ respectively. In the case of model II, more favorable results are observed when the variables $n$ increases and $\alpha$ decreases. The numeric values of $\Gamma$ at the center are $1.33$ and $1.40$ for the values of $\alpha=-0.1,-0.5$, respectively, where we fixed $n=0.6$. Through this investigation, we have evaluated that the decoupling parameter $\alpha$ is constrained within the limit of $-4.1<\alpha<0.2$ for Model I and $-\infty < \alpha < 0.8$ for Model II.

\subsection{Velocity of sound via cracking method}

Generating a physically accurate model, it requires verification of the causality criterion, which specifies that the speed of sound inside the compact object must be subluminal. The following equation can be used to compute the sound velocity of the stellar fluid.

\begin{eqnarray}
     v_r^2=\frac{dp_r^{\text{eff}}}{d\rho^{\text{eff}}}\, ,\quad
v_t^2=\frac{dp_t^{\text{eff}}}{d\rho^{\text{eff}}}.
\end{eqnarray} 

In a comprehensive set of lectures \cite{55,56,57}, Herrera et al. extensively investigated the concept of stellar structure cracking, focusing on incorporating anisotropic matter structures. The concept of cracking, or overturning, was initially proposed in 1992. This approach becomes advantageous in detecting and characterizing potentially unstable anisotropic matter formations. From Figs. \ref{sound} and \ref{sound2}, one can see that the speed of sound maintains the inequality $0<v_r^2<1,\,0<v_t^2<1$, which meets the causality criterion.
\\
 \begin{table*}[!htp]
    \centering
        \caption{The corresponding numerical values of the central density ($\rho_c$), surface density ($\rho_s$), central pressure ($p_c$), charge at surface ($q_s$), and central value of adiabatic index ($\Gamma_c$) for different values of the model parameter $n$ with $R=10.00\, \text{km}$, $A=0.01$.}\label{table1}
            \begin{tabular}{@{}ccccccccccccc@{}}
            \hline
            $\alpha$ & $n$ & Cases & $\rho_c$  & $\rho_s$ &  $p_c$ & $q_s$& $\Gamma_c$\\
            & & & $\text{gm}/\text{cm}^3$ & $\text{gm}/\text{cm}^3$ & $\text{dyne}/\text{cm}^2$& $\text{coulomb(C)}$ & &\\
            \hline\hline
            & & & & Model-I& & & &\\
            0.0 & $0.6$ & $f(Q,T)$ & $8.51732\times 10^{13}$ & $2.56654\times 10^{13}$ &$2.1879\times 10^{33}$ & $1.11776\times 10^{20}$ & $1.37$\\
            0.0 & $0.8$ & $f(Q,T)$ & $7.36505\times 10^{13}$ & $2.20918\times 10^{13}$ &$2.09554\times 10^{33}$ & $1.136\times 10^{20}$ & $1.46$\\
            0.2 & $0.6$ & $f(Q,T)+\text{MGD}$ & $8.66316\times 10^{13}$ & $2.54219\times 10^{13}$ &$1.75037\times 10^{33}$ & $1.11776\times 10^{20}$ & $1.33$ \\
            0.2 & $0.8$ & $f(Q,T)+\text{MGD}$ & $7.50473\times 10^{13}$ & $2.18696\times 10^{13}$ &$1.67643\times 10^{33}$ & $1.136\times 10^{20}$ & $1.40$ \\
            \hline\hline
            & & & & Model-II& & & &\\
             0.0 & $0.6$ & $f(Q,T)$ & $8.51732\times 10^{13}$ & $2.56654\times 10^{13}$ &$1.71605\times 10^{18}$ & $1.11776\times 10^{20}$ &$1.37$\\
            0.0 & $0.8$ & $f(Q,T)$ & $7.36505\times 10^{13}$ & $2.20918\times 10^{13}$ &$1.20931\times 10^{18}$ & $1.136\times 10^{20}$ & $1.46$\\
            -0.2 & $0.6$ & $f(Q,T)+\text{MGD}$ & $6.8333\times 10^{13}$ & $2.06529\times 10^{13}$ &$2.98541\times 10^{33}$ & $1.10263\times 10^{20}$ & $1.41$ \\
            -0.2 & $0.8$ & $f(Q,T)+\text{MGD}$ & $5.91043\times 10^{13}$ & $1.77834\times 10^{13}$ &$2.71892\times 10^{33}$ & $1.12157\times 10^{20}$ & $1.47$ \\
            \hline
        \end{tabular}
    \end{table*}

\begin{figure*}[!htp]
\includegraphics[width=8cm, height=6cm]{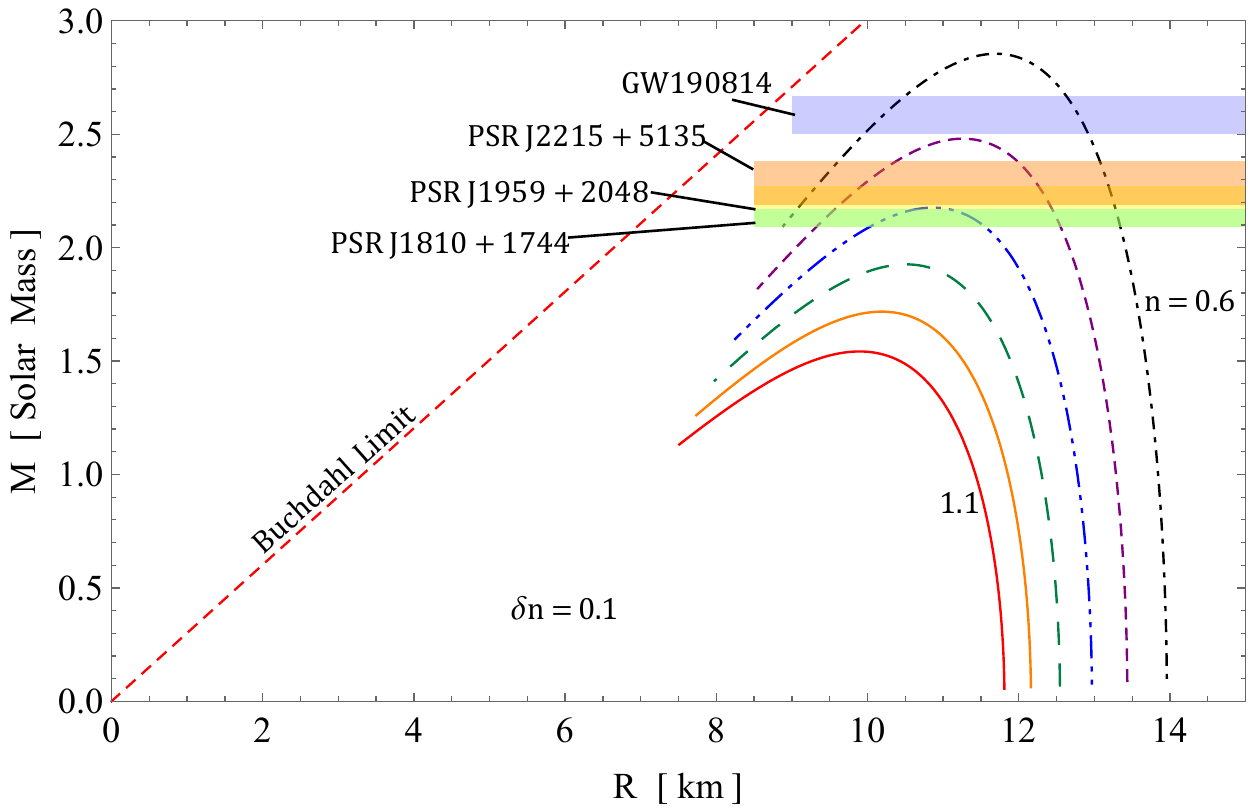}\,\,\,\,\,\,\,
\includegraphics[width=8cm, height=6cm]{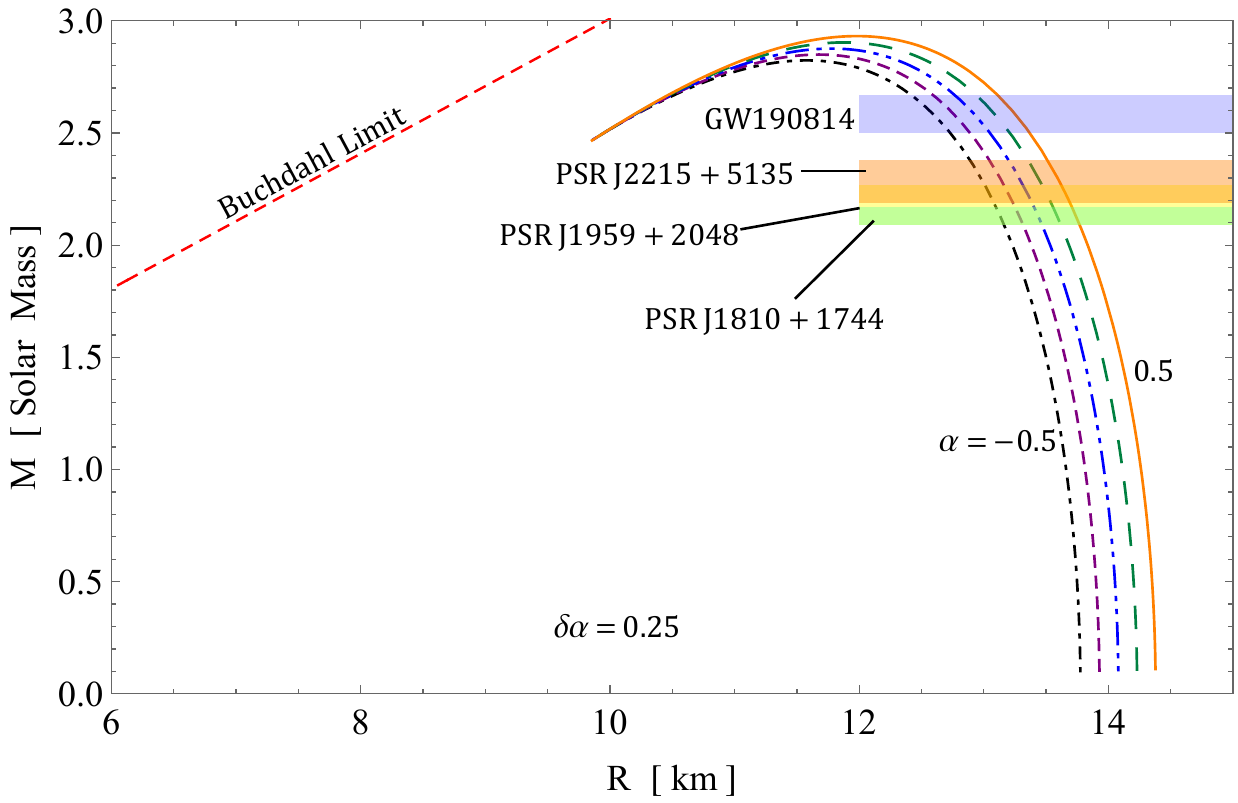}
\caption{$M-R$ curves for model-I. Here $\alpha=-0.2,  A=0.01$ with $m = -0.1$ (Left) and $n = 0.6$ (Right).  \label{MR1}}
\end{figure*}

\begin{figure*}[!htp]
\includegraphics[width=8cm, height=6cm]{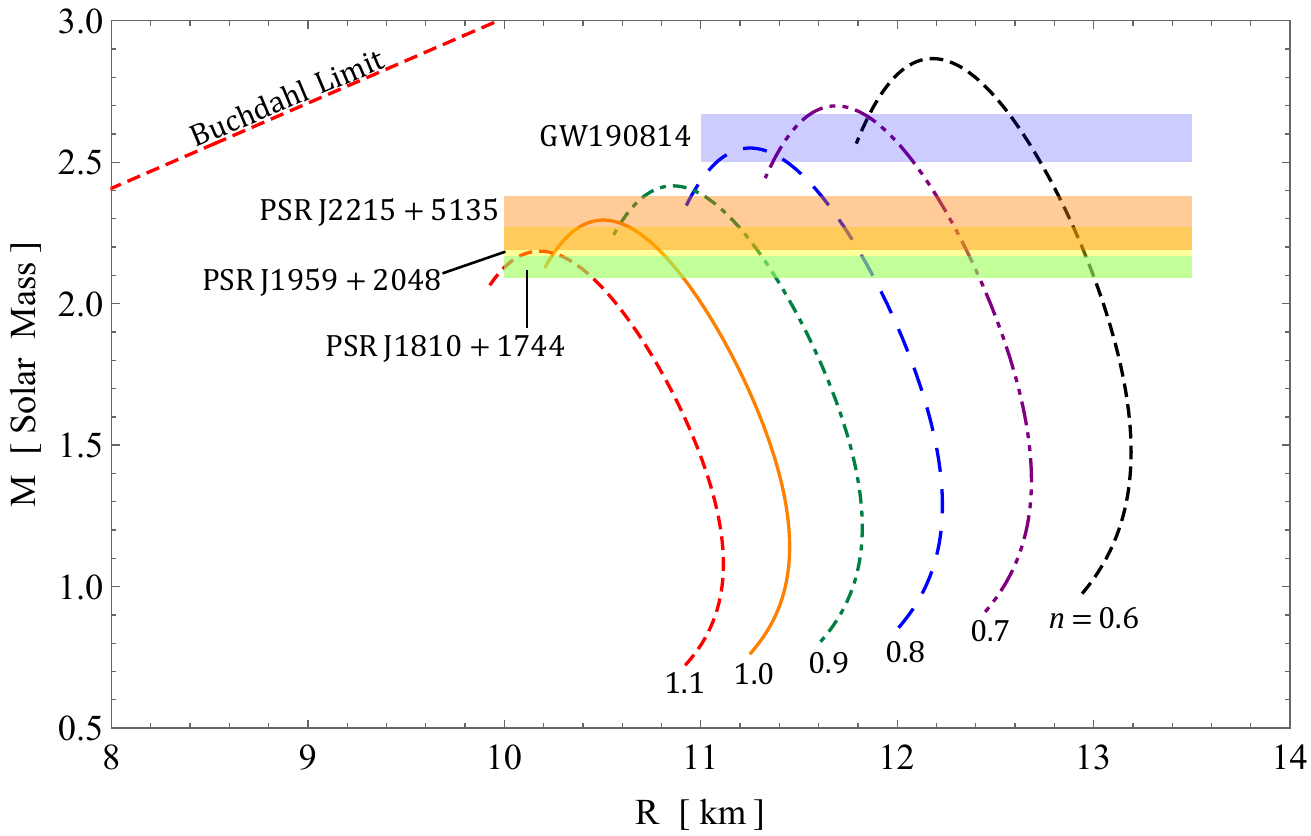}\,\,\,\,\,\,\,
\includegraphics[width=8cm, height=6cm]{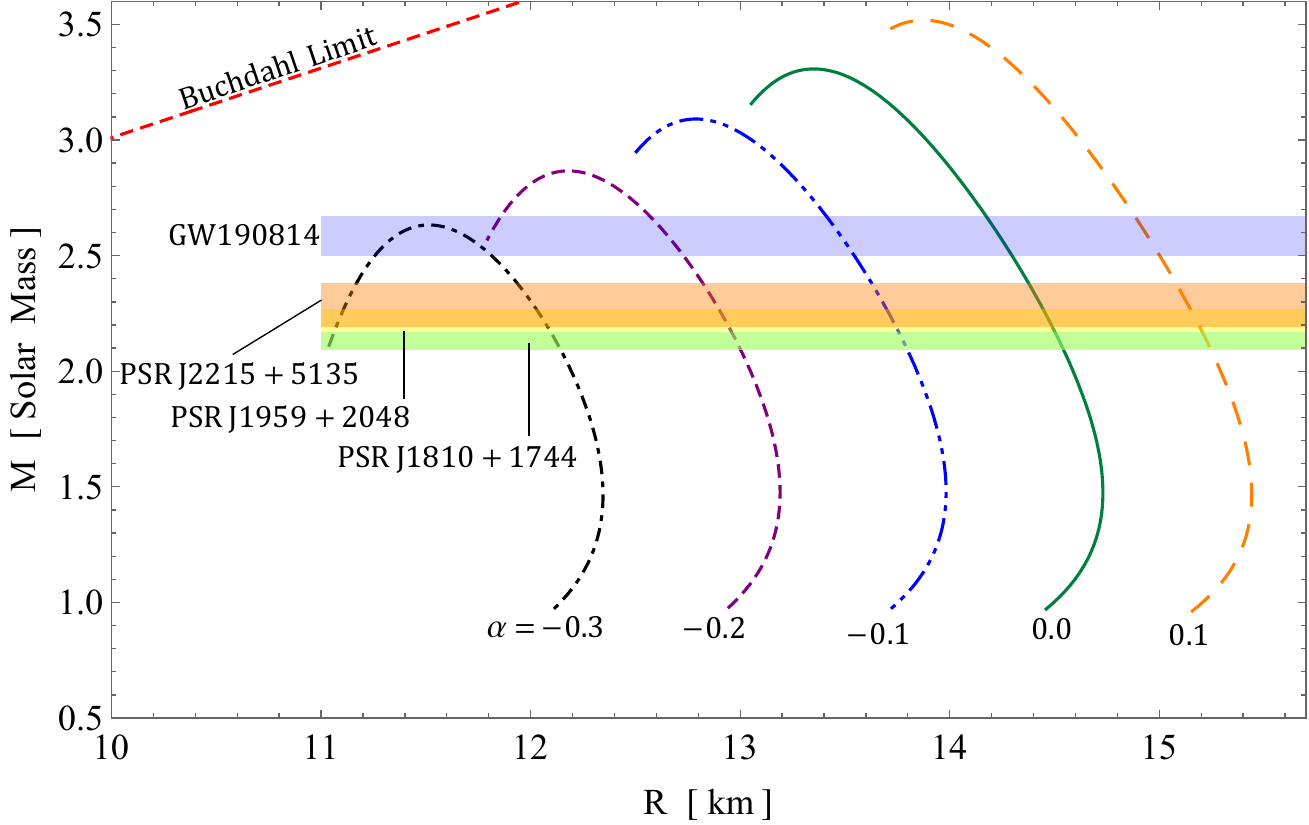}
\caption{$M-R$ curves for for model-II. Here $\alpha=-0.2,  A=0.01$ with $m = -0.035$ (Left) and $n = 0.6$ (Right).  \label{MR2}}
\end{figure*}
\begin{table*}[!htp]
\centering
\caption{The predicted radii of few high mass compact stars  corresponding Fig. \ref{MR1}.}\label{table6}
 \scalebox{0.85}{\begin{tabular}{ |*{10}{c|}}
\hline
 &    & \multicolumn{3}{c|}{{Predicted $R$ km}} & \multicolumn{5}{c|}{{Predicted $R$ km}}  \\[0.15cm]
\cline{3-10}
{Objects} & {$\frac{M}{M_\odot}$} & \multicolumn{3}{c|}{ $n$} & \multicolumn{5}{c|}{$\alpha$} \\[0.15cm]
\cline{3-10}
&  & $0.6$ & $0.7$ & $0.8$ & $-0.5$ & $-0.25$ & $0.0$ & $0.25$ & $0.50$\\[0.15cm] \hline
PSR J1810+1744 \citep{r1-PSRJ1810+1744,r1-PSRJ1810+1744-1} & 2.13$\pm$0.04  & $13.31_{-0.03}^{+0.04}$  &  $12.50_{-0.05}^{+0.06}$  &   $11.38_{-0.26}^{+0.20}$  &  $13.14_{-0.05}^{+0.02}$ &  $13.28_{-0.02}^{+0.04}$ & $13.44_{-0.03}^{+0.02}$ & $13.58_{-0.03}^{+0.04}$ & $13.74_{-0.04}^{+0.03}$\\[0.15cm]
\hline
PSR J1959+2048 \citep{star1} & 2.18$\pm$0.09  & $13.27_{-0.01}^{+0.01}$  &  $12.42_{-0.03}^{+0.03}$  &   -  &  $13.09_{-0.04}^{+0.01}$ &  $13.23_{-0.01}^{+0.02}$ & $13.39_{-0.02}^{+0.02}$ & $13.54_{-0.01}^{+0.01}$ & $13.70_{-0.01}^{+0.01}$ \\[0.15cm]
\hline
PSR J2215+5135 \citep{star1} & $2.28^{+0.10}_{-0.09}$ & $13.19_{-0.12}^{+0.07}$  &  $12.25_{-0.27}^{+0.16}$  & -  &  $13.00_{-0.11}^{+0.07}$ &  $13.15_{-0.10}^{+0.08}$ & $13.30_{-0.10}^{+0.08}$ & $13.45_{-0.08}^{+0.09}$ & $13.60_{-0.09}^{+0.09}$ \\[0.15cm]
\hline
GW190814 \citep{waves3} & 2.5-2.67 & $12.82_{-0.18}^{+0.09}$  &  -  &   -  &  $12.61_{-0.18}^{+0.12}$ &  $12.78_{-0.18}^{+0.11}$ & $12.93_{-0.16}^{+0.11}$ & $13.10_{-0.14}^{+0.10}$ & $13.28_{-0.16}^{+0.09}$\\[0.15cm]
\hline
\end{tabular}}
\end{table*}
\begin{table*}[!htp]
\centering

\caption{The predicted radii of few high mass compact stars  corresponding Fig. \ref{MR2}.}\label{table7}
 \scalebox{0.67}{\begin{tabular}{ |*{13}{c|}}
\hline
 &    & \multicolumn{6}{c|}{{Predicted $R$ km}} & \multicolumn{5}{c|}{{Predicted $R$ km}}  \\[0.15cm]
\cline{3-13}
{Objects} & {$\frac{M}{M_\odot}$} & \multicolumn{6}{c|}{ $n$} & \multicolumn{5}{c|}{$\alpha$} \\[0.15cm]
\cline{3-13}
&  & $0.6$ & $0.7$ & $0.8$ & 0.9 & 1.0 & 1.1 & $-0.3$ & $-0.25$ & $-0.1$ & $0.0$ & $0.1$\\[0.15cm] \hline
PSR J1810+1744 \citep{r1-PSRJ1810+1744,r1-PSRJ1810+1744-1} & 2.13$\pm$0.04  & $12.98_{-0.02}^{+0.02}$  &  $12.39_{-0.02}^{+0.03}$  &   $11.85_{-0.03}^{+0.03}$  &  $11.35_{-0.04}^{+0.03}$ &  $10.86_{-0.05}^{+0.05}$ & $10.39_{-0.10}^{+0.05}$ & $12.12_{-0.01}^{+0.02}$ & $12.98_{-0.01}^{+0.02}$ & $13.78_{-0.02}^{+0.02}$ & $14.52_{-0.02}^{+0.02}$ & $15.22_{-0.02}^{+0.03}$\\[0.15cm]
\hline
PSR J1959+2048 \citep{star1} & 2.18$\pm$0.09  & $12.95_{-0.01}^{+0.01}$  &  $12.36_{-0.01}^{+0.01}$  &   $11.81_{-0.01}^{+0.01}$  &  $11.30_{-0.02}^{+0.01}$ &  $10.80_{-0.02}^{+0.03}$ & $10.25_{-0.05}^{+0.07}$ & $12.09_{-0.01}^{+0.01}$ & $12.95_{-0.01}^{+0.02}$ & $13.75_{-0.02}^{+0.01}$ & $14.49_{-0.01}^{+0.01}$ & $15.20_{-0.01}^{+0.01}$ \\[0.15cm]
\hline
PSR J2215+5135 \citep{star1} & $2.28^{+0.10}_{-0.09}$ & $12.89_{-0.07}^{+0.05}$  &  $12.28_{-0.08}^{+0.06}$  &   $11.72_{-0.10}^{+0.07}$  &  $11.18_{-0.15}^{+0.09}$ &  $10.61_{-}^{+0.16}$ & - & $12.02_{-0.08}^{+0.05}$ & $12.89_{-0.07}^{+0.06}$ & $13.69_{-0.06}^{+0.05}$ & $14.43_{-0.05}^{+0.05}$ & $15.14_{-0.06}^{+0.05}$ \\[0.15cm]
\hline
GW190814 \citep{waves3} & 2.5-2.67 & $12.66_{-0.09}^{+0.06}$  &  $12.00_{-0.14}^{+0.08}$  &   -  &  - &  - & - & $11.71_{-}^{+0.10}$ & $12.67_{-0.10}^{+0.05}$ & $13.49_{-0.08}^{+0.05}$ & $14.24_{-0.06}^{+0.05}$ & $14.95_{-0.06}^{+0.05}$\\[0.15cm]
\hline
\end{tabular}}
\end{table*}

\section{Observations of the Maximum Mass Limit for Strange Stars Using M-R Diagrams}\label{sec:XA}

We have displayed the mass profile as a function of radius for the $\theta_1^1=p_r$ sector in Fig.~\ref{MR1} by varying the model parameter $n$ (left panel) and $\alpha$(right panel). One can observe that for both the panels, these models are able to explain the existence of compact objects with masses in the range of $2.0\,\text{M}_{\odot}-2.8\,\text{M}_{\odot}$. The fact that the observed masses are higher than the generally accepted figures for neutron stars makes this observation significant. 
The upper mass profiles effectively correspond to the objects involved in the GW190814 event \cite{gw1}. GW190814 departed from the initial groundbreaking gravitational wave events, as it differed by not resulting from two colliding black holes. Instead, the ripples in the fabric of space-time during this event originated from the merger of two neutron stars having mass $2.7\, M_{\odot}$. 
In this context, we want to highlight that the secondary mass of GW190814 falls within the proposed lower "mass gap" of 2.5 to 5 $M_{\odot}$, \cite{MRC1,MRC2} situated between known neutron stars and black holes which align in the recent study of \cite{PB1} in $f(Q)$ gravity. It is evident from the figure that for the higher value of the decoupling parameter $\alpha$, the mass limit increases, and for both panels, one can observe that the observable mass does not exceed the Buchdahl limit. Therefore, for a fixed model parameter $n$, the higher value of $\alpha$ leads to the maximum mass limit for a compact object. On the other hand, for a fixed $\alpha$, and the lower value of model parameter $n$, our model-I satisfies the higher mass profiles for a massive compact object. To draw a comparison with $f(Q)$ gravity, it is noteworthy that Bhar and Pretel's research \cite{PB2} has shown an increase in both the maximum mass and the corresponding radius of dark energy stars as the coupling parameter in $f(Q)$ gravity increases. Contrarily, in $f(Q,T)$ gravity, the results indicate that as the coupling parameter between non-metricity and matter decreases, the maximum mass limit for a compact object is achieved. Again In quadratic $f(Q)$ gravity, when a quintessence field is present, the maximum mass and corresponding radius increase as the value of the coupling parameter of $f(Q)$ gravity decreases \cite{PB3}. Additionally, other studies have conducted a detailed analysis of the maximum allowable mass in compact objects using the $M-R$ curve in the context of modified gravity\cite{SD2,SD1}. The mass profile for the solution $\Theta_0^0=\rho(r)$ has been displayed in Fig.~\ref{MR2}. We have varied the model parameter $n$ (by fixing $\alpha$ in the left panel) and the decoupling parameter $\alpha$ (by fixing $n$ in the right panel) in a wide range.

Subsequently, it becomes evident that an elevation in the decoupling parameter $\alpha$ leads to a substantial rise in the mass of the stellar configuration. It is worth mentioning that the upper mass limit for a varying $\alpha$ parameter exceeds $3 {M_{\odot}}$. Furthermore, we have presented the predicted radii of some massive compact objects corresponding to the solutions $\Theta_1^1=p_r$ and $\Theta_0^0=\rho_r$ in the tables \ref{table6} and \ref{table7}, respectively.
In Figs.~\ref{MR1} and \ref{MR2}, we displayed some well-fitted compact stars based on our model in the mass-radius plot. The best-fitted compact stars are, GW190814 (mass $2.7 M_{\odot}$), PSR J2215+5135 (mass = $2.13\pm 0.04 M_{\odot}$), PSR J1959+2048 (mass = $2.18 \pm 0.09 M_{\odot}$), PSR J1810+1744 (mass = $2.13\pm 0.04 M_{\odot}$). For our model I, we can see the maximum allowable mass for model-I is:  $2.87 M_{\odot}$ with the radius $11.7$ km for $n=0.6$ while $2.95 M_{\odot}$ with the radius $11.7$ km the radius $12$ km for $\alpha=0.5$. However, model-II possess the maximum mass $2.95 M_{\odot}$ with radius $12.2$ km for $n=0.6$ and while mass $3.6 M_{\odot}$ with the radius $13.9$ km for $\alpha=0.1$.


\section{Concluding Remarks}\label{XII}
This article has studied a self-gravitating strange star model with the Class I relativistic solution within the $f(Q, T)$ gravity theory framework. An appropriate form of the function $g(r)$ is imposed in accordance with the notion of MGD, which guarantees the regularity of the metric potentials and all physical characteristics of the system within the stellar configuration. The aforementioned technique offers a methodical approach in the following manner: (i) Firstly, the main decoupling system is divided into two subsystems. (ii) Secondly, The resolution of these subsystems in an independent manner. For obtaining the seed solution i.e., the set of equations with $\alpha=0$, we have utilized the embedding Class I condition. We have obtained a class of exact solution for the $\Theta$ sector by mimicking the pressure constraint as $\Theta_1^{1}=p(r)$ and $\Theta_0^{0}=\rho(r)$ for the model I and model II respectively. In the next step, we applied the continuity of the first and second fundamental forms to determine the constant in our proposed model. The findings of the current investigation are typically interesting, unique, and acceptable regarding their physical feasibility. The following can be highlighted as certain significant aspects of the current investigation: 
\begin{itemize}
        \item  Based on the graphical analysis presented in Figs. \ref{hqr} and \ref{hqr2}, it can be observed that the electric charge within the star exhibits a zero value at its center while progressively growing towards the star's surface. This observation aligns with a fundamental finding in astrophysics, which states that all charges must be concentrated at the surface region of the celestial object. Moreover, it is evident that the electric charge increases when the variable $n$ grows. It should be noted that the influence of the parameter $\alpha$ on the electric charge is nonexistent for model I. We can conclude through the graphical analysis of the deformation function in Fig.~\ref{hqr}, which demonstrates that the function $h(r)$ becomes zero at the boundary for all values of the model parameter $n$ hence not influencing the total mass of the star for model-I. In contrast, in model II, the deformation function exhibits a progressive trend towards the boundary of the stars; hence, in this case, the total mass would be the $f(Q,T)$ mass in addition to MGD mass.
        
         \item  From Figs.~\ref{pt} and \ref{pt2}, one can observe that all the physical quantities remain positive and give finite values throughout the stellar region for both of our constructed models.  In addition, the effective density, radial pressure, and tangential pressure exhibit a peak near the central region of the star, followed by a progressive decline towards the outer radius. In regard to the anisotropy component, it can be asserted that our proposed model yields superior outcomes compared to the $f(Q,T)$ model. Effective anisotropy is a measure of the disparity between the pressures applied in the tangential and radial directions, therefore, it demands to be unequal. However, when considering the $f(Q,T)$ model, readers can observe that the $\Delta$ graph remains constant, indicating zero value throughout the stellar configuration. For this reason, it can be asserted that the MGD technique is much more beneficial for describing the strange star model in $f(Q,T)$ gravity.
         
         \item Another essential parameter for characterizing the relationship between matter density and pressure is by the determination of the equation of state. The graphical representation of the equation of state can be observed in Figs. \ref{eos} and \ref{eos2}. The findings of our study suggest that both of these physical parameters exhibit a peak value in the vicinity of the star's central region, gradually decreasing as one moves toward the outer regions of the star. Moreover, it should be noted that the quantities $\omega_r$ and $\omega_t$ fall within the range of the radiation era, specifically satisfying the condition $0\leq \omega_r, \, \omega_t \leq 1$. 

        \item  From the Fig. \ref{mass2}, it is observable that the surface redshift is monotonically growing, but it lies on $0\leq \mathcal{Z_{\mathrm{s}}}\leq 1$.

        \item A comparative study of the adiabatic index with respect to the radial coordinate '$r$' is presented graphically in Fig.~\ref{adb2}. The comparison is conducted among the cases $f(Q,T)$, and $f(Q,T)$+MGD. Based on the depicted image, it is observed that our developed decoupled strange star model provides a more precise and feasible representation for both models, as the adiabatic index consistently exceeds the threshold of $4/3$. 
            \item The speed of sound was visually examined using Herrera's cracking concept to conduct a stability analysis of our suggested model. It has been found in Figs.~\ref{sound} and \ref{sound2} that in all situations, the radial and tangential components of the speed of sound fall within the stability range, thereby confirming the stability of our model.
         \end{itemize} 

 Finally, the mass and the corresponding radii for the pulsars PSR J1810+174, PSR J1959+2048, PSR J2215+5135, and GW190814 are examined for both models in the framework of $f(Q,T)$ gravity. In Figs. \ref{MR1} and \ref{MR2} along with Tables \ref{table6} and \ref{table7}, a comparison is made between the predictions made by the models and the characteristics that have been observed for these compact stars. The observed data shows a good agreement with the observational data.    
 
Based on the comprehensive analysis conducted above,  it can be asserted that our suggested model in $f(Q,T)$ gravity, utilizing the MGD technique, demonstrates great acceptability and usefulness for studying the compact star models. \\

\textbf{Data availability:} There are no new data associated with this article.

\section*{Acknowledgements}
SP \& PKS  acknowledges the National Board for Higher Mathematics (NBHM) under the Department of Atomic Energy (DAE), Govt. of India for financial support to carry out the Research project No.: 02011/3/2022 NBHM(R.P.)/R \& D II/2152 Dt.14.02.2022. 
\newpage

\begin{widetext}
\section*{Appendix-I}
    \begin{eqnarray}
      \Theta_0^{0} &=& \frac{1}{2 (n+1) (2 n+1) (A r^2+1)^3 (B+5 C r^2)^2} \Bigg[m \big(A^3 (n+1) r^4 \big(B^2-2 B C r^2+5 C^2 r^4\big)+A^2 r^2 \big(B^2 (5 n+3)+12 B C (n-1) r^2 \nonumber \\&& +15 C^2 (5 n+3) r^4\big)+6 A B^2-2 A B C (19 n+1) r^2+30 A C^2 (n+2) r^4-4 C (3 n+2) (3 B+5 C r^2)\big)\Bigg],\nonumber\\
     \Theta_2^{2} &=& \frac{m}{(n+1)^2 (2 n+1)^2 (A r^2+1)^5 (B+C r^2) (B+5 C r^2)^3} \Bigg[\Big(-A^5 C (n+1)^2 r^{10} (B+C r^2) \big(B^2 (2 n+1)+2 B C r^2\nonumber\\&&-25 C^2 (2 n+1) r^4\big)+A^4 (n+1) r^6 \big(B^4 n (2 n+1)+B^3 C (18 n^2-5) r^2+B^2 C^2 (n (116 n+71)-6) r^4\nonumber\\&&+B C^3 (n (370 n+549)+172) r^6-5 C^4 (2 n+1) (5 n-13) r^8\big)+A^3 r^4 \big(B^4 (n+1) (2 n+1)^2\nonumber\\&&+B^3 C (n (n (34 n+53)+23)+2) r^2+B^2 C^2 \big(n (216 n^2+418 n+271)+43\big) r^4+5 B C^3 (n (2 n (53 n+150)\nonumber\\&&+301)+89) r^6-5 C^4 (n (n (200 n+403)+216)+31) r^8\big)+A^2 r^2 \big(B^4 (n+1) (n+2) (2 n+1)+B^3 C (3 n (n (3-2 n)\nonumber\\&&+8)+11) r^2+2 B^2 C^2 (n (2 n (18-5 n)+67)+26) r^4+B C^3 (n (n (211-150 n)+716)+329) r^6-5 C^4 (n (9 n (50 n+113)\nonumber\\&&+697)+148) r^8\big)+A \big(B^4 (n+1) (2 n+1)-16 B^3 C n (n+1) (2 n+1) r^2-B^2 C^2 (n (n (214 n+363)+194)+37) r^4\nonumber\\&&- 2 B C^3 (3 n (n (110 n+177)+84)+37) r^6-5 C^4 (n (n (390 n+889)+653)+158) r^8\big)\nonumber\\&&-2 C (3 n+2) \big(B^3 (n+1) (2 n+1)+8 B^2 C (n+1) (2 n+1) r^2+B C^2 (n (50 n+63)+17) r^4+50 C^3 (n+1) (2 n+1) r^6\big)\Big)\Bigg],
     \nonumber
    \end{eqnarray}
    
\section*{Appendix-II}
    \begin{eqnarray}
       \Theta_{1}^{1} &=& -\frac{A m^2 (B+5 C r^2) (A B (7 n+5)-C (9 n+7))}{4 (2 n^2+3 n+1) (A r^2+1) (A B-C) (B+Cr^2)}, \nonumber\\
        \Theta_2^{2} &=& \frac{-1}{8 (n+1)^2 (2 n+1)^2 (A r^2+1)^3 (C-A B)^2 (B+C r^2)} \Bigg[A m^2 (A B (7 n+5)-C (9 n+7)) \big(4 A^2 C (n+1) (2 n+1) r^6 (A B-C) \nonumber\\&&+A C r^4  (-A B m (7 n+5)+10 A B (n+1) (2 n+1)+C m (9 n+7)-10 C (n+1) (2 n+1))\nonumber\\&&+2 (n+1) (2 n+1) r^2 (A B-C) (A B+3 C)+2 B (n+1) (2 n+1) (A B-C)\big)\Bigg].\nonumber
    \end{eqnarray}
\end{widetext}

\end{document}